\documentclass[aps,prd,a4paper,nofootinbib]{revtex4-2}

\usepackage{graphicx}
\usepackage{bm}
\usepackage{amsfonts}
\usepackage{amsmath}
\usepackage{float}
\usepackage{rotating}
\usepackage{xspace}
\usepackage[breaklinks=true]{hyperref}
\hypersetup{colorlinks=true,linkcolor=black,citecolor=black,urlcolor=black}

\usepackage[export]{adjustbox}
\usepackage{tikz}
\usetikzlibrary{matrix}

\usepackage[english]{babel}

\newcommand{\ve}[1]{\mbox{\boldmath$#1$}}
\let\vec=\ve

\def\vx{{\ve x}}

\providecommand\gaia{\textit{Gaia}\xspace}
\providecommand\gaiaNIR{\textit{Gaia}\textrm{NIR}\xspace}

\newcommand*{\hPlusSin}{h_{+}^{\mathrm{s}}}
\newcommand*{\hPlusCos}{h_{+}^{\mathrm{c}}}
\newcommand*{\hTimesSin}{h_{\times}^{\mathrm{s}}}
\newcommand*{\hTimesCos}{h_{\times}^{\mathrm{c}}}

\newcommand*{\hps}{\hPlusSin}
\newcommand*{\hpc}{\hPlusCos}
\newcommand*{\hts}{\hTimesSin}
\newcommand*{\htc}{\hTimesCos}

\newcommand{\lbar}{\overline{l}}
\newcommand{\cE}{{\cal E}}

\newcommand{\orcit}[1]{\protect\href{https://orcid.org/#1}{\protect\includegraphics[width=8pt]{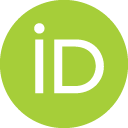}}}

\let\oldbibitem\bibitem
\renewcommand\bibitem[2][]{\oldbibitem{#2}}

\makeatletter
\renewcommand*\env@matrix[1][\arraystretch]{\edef\arraystretch{#1}\hskip -\arraycolsep
	\let\@ifnextchar\new@ifnextchar
	\array{*\c@MaxMatrixCols c}}
\makeatother

\begin{document}

\title{Gravitational waves and small-field astrometry}

\author{Robin Geyer\,\orcit{0000-0001-6967-8707}}
\email{Contact author: robin.geyer@tu-dresden.de}
\author{Sven Zschocke\,\orcit{0000-0002-2699-1063}, Michael Soffel, Sergei Klioner\,\orcit{0000-0003-4682-7831}}

\affiliation{Lohrmann Observatory, Technische Universität Dresden,
              Helmholtzstraße 10, 01069 Dresden, Germany}

\author{Lennart Lindegren\,\orcit{0000-0002-5443-3026}}

\affiliation{Lund Observatory, Division of Astrophysics, Department of
  Physics, Lund University, Box 43, 22100 Lund, Sweden}

\author{Uwe Lammers\,\orcit{0000-0001-8309-3801}}

\affiliation{European Space Agency (ESA), European Space Astronomy Centre
              (ESAC), Camino bajo del Castillo, s/n, Urbanización Villafranca del
              Castillo, Villanueva de la Cañada, 28692 Madrid, Spain}

\begin{abstract}
Astrometric observations can, in principle, be used to detect
gravitational waves.
In this paper we give a practical overview of the gravitational
wave effects which can be expected specifically in small-field astrometric data.
Particular emphasis is placed on the differential effect between pairs
of sources within a finite field of view.
We also present several general findings that are not restricted to the small-field case.
A detailed theoretical derivation of the general astrometric effect of a
plane gravitational wave is provided.
Numerical simulations, which underline our theoretical
findings, are presented.

We find that small-field missions suffer from significant detrimental
properties, largely because their relatively small fields only allow
the measurement of small differential effects which can be expected
to be almost totally absorbed by standard plate calibrations.
\end{abstract}

\maketitle

\begin{center}
\today
\end{center}

\section{Introduction}
\label{Introduction}

Over the past several years, we have witnessed a growing interest in designing
space-based small-field astrometric telescopes of ultimate sub-$\mu$as
accuracy and their possible scientific applications (see e.g. \cite{2021ExA....51..845M,2022SPIE12180E..1FM,2024PASJ...76..386K}).
Moreover, the idea has emerged that even non-astrometric, imaging space
telescopes could be used to make high-precision astrometric
measurements \cite{2023MNRAS.525.2585N,2024A&A...692A..96L}.
On the other hand, space-based global astrometry, like \gaia, has already shown that it is able to deliver a remarkable observational accuracy
\cite{2021A&A...649A...2L,2022A&A...667A.148G,2021A&A...649A...9G}.
At the same time, with gravitational wave (GW) astronomy now a routine
reality \cite{2016PhRvL.116f1102A,2025arXiv250818080T,2025MNRAS.536.1501G}, there is a
certain interest in also using astrometric measurements to detect GW
signals
\cite{Article_Book_Flanagan,2017PhRvL.119z1102M,2018PhRvD..97l4058M,2018PhRvD..98b4020O}.
The effects of GWs on astrometric measurements have been known for a 
long time \cite{1990NCimB.105.1141B,1996ApJ...465..566P,1997ApJ...485...87G}.

While it has already been discussed in the literature that
all-sky astrometric missions like \gaia\ are, in principle, sensitive to GWs
\cite{2018CQGra..35d5005K,2025A&A...695A.172G}, a practical yet
succinct discussion for small-field missions is still lacking.

In this paper, we present a practical overview of the GW effects
that can be expected in small-field astrometry. In
Section~\ref{sec__max_change}, we discuss various upper estimates of
the astrometric GW effect both for global astrometry and for
small-field astrometry with a given field of view (FoV) size. Section~\ref{section-numerical} contains the results from
numerical simulations which validate and illustrate the theoretical formulas.
There, we also discuss the appearance of the astrometric GW effects in a given
small FoV. A concluding discussion can be found in Section
\ref{sec__conclusion}. Appendix~\ref{sec__appendix} contains a concise
theoretical derivation of the astrometric GW effect from the basic
principles. Several important aspects of the effect are elucidated there.

In the whole paper, we will discuss GWs with an effectively plane wavefront
at the observer and at the observed astrometric source.
For all simulations, and discussions of the maximum measurable signal, we use Eq.~\eqref{proof_tetrad_observer_i}, or, if angles between sources are concerned, Eq.~\eqref{final_result_angle}.
Finally, a note with respect to terminology: for the rest of the paper,
we will refer to the source of
the gravitational wave itself as ``GW emitter(s)'', never as
``source(s)''. Conversely, the sources which we astrometrically observe
we refer to as ``astrometric source(s)'' or just ``source(s)''.
 
\section{Maximal change of the angular distance between sources due to a GW}
\label{sec__max_change}

In this Section, we use the theoretical formulation of the astrometric
GW effect given in Appendix~A of \cite{2025A&A...695A.172G} to compute
theoretical upper estimates of the magnitude of the astrometric GW
effect in various situations.
Appendix~A.1 of \cite{2025A&A...695A.172G} gives the basic formulas for the effect,
while Appendix~A.2 gives an alternative representation of the formulas
given in Appendix~A.1, which is especially useful for the astrometric discussion of the effect. Appendix~\ref{sec__appendix} of this work
contains a basic derivation of the formulas from Appendix~A.1 of
\cite{2025A&A...695A.172G}.

Equation~(A.11) of \cite{2025A&A...695A.172G} gives an expression for
the angular variation of the observed position of a source due to a
GW. The components of the two-dimensional displacement vector
$(\delta\alpha^*,\delta\delta)$ are given relative to the local
coordinate system on the sphere attached to the undisturbed position
of the source, represented by the local dyad
$(\vec{e}_\alpha,\vec{e}_\delta)$.

One can show that the maximal change of the source position
\begin{equation}
	\label{estimate1}
	\sqrt{(\delta\alpha^*)^2+(\delta\delta)^2}\le \Delta_{\rm max}\,\sin\theta,
\end{equation}
where $\theta$ is the angle between the source with coordinates
$(\alpha,\delta)$ and the direction of GW propagation $(\alpha_{\rm
  GW},\delta_{\rm GW})$, $\Delta_{\rm max} = {(2-e^2)}^{-1/2}\,h/2$ is
the maximal astrometric effect, and
$h={\left({\left(\hpc\right)}^2+{\left(\hps\right)}^2+{\left(\htc\right)}^2+{\left(\hts\right)}^2\right)}^{1/2}$
is given by the strain and phase parameters of the GW as in
Eqs.~(A.19)--(A.21) of \cite{2025A&A...695A.172G}---see the reasoning
in Appendix~A.2 of \cite{2025A&A...695A.172G}.  Here, $e$ is the
eccentricity of the elliptic astrometric signal caused by the GW.
This parameter is defined by Eq.~(A.17) of \cite{2025A&A...695A.172G}
and depends only on the strain and phase parameters of the GW.  We note
that $0\le\theta\le\pi$ and $0\le\sin\theta\le1$. In all equations
here, we consider a linear model for the GW effect, so that terms
${\cal O}(\Delta_{\rm max}^2)$ are always neglected.

The general discussion in \cite{2025A&A...695A.172G} is relevant for
absolute astrometric observations like those of \gaia. However, for
any kind of differential observations---that is, observations of the
angular distances between pairs of sources---this discussion
immediately gives an upper estimate of observable variations of the
angular distance due to a GW. We denote the undisturbed observable
angular distance between source A and source B as
$\psi_\mathrm{AB}$. This is the angular distance in the absence of the
GW and can be computed from the unperturbed positions
$(\alpha_i,\delta_i)$, $i = \mathrm{A},\mathrm{B}$, e.g. as
$\cos\psi_\mathrm{AB}=\sin\delta_\mathrm{A}\sin\delta_\mathrm{B}+\cos\delta_\mathrm{A}\cos\delta_\mathrm{B}\cos(\alpha_\mathrm{B}-\alpha_\mathrm{A})$.
We denote the disturbed observable angular distance between these sources in the
presence of the GW as $\psi_\mathrm{AB}^{\rm gw}$. The observable
change of the angular distance,
$\delta\psi_\mathrm{AB}=\psi_\mathrm{AB}^{\rm gw}-\psi_\mathrm{AB}$,
can obviously be estimated as
\begin{equation}
  \label{estimate2}
|\delta\psi_\mathrm{AB}|\le \Delta_{\rm max}\,(\sin\theta_\mathrm{A}+\sin\theta_\mathrm{B})
\end{equation}
for a given GW with a known maximal astrometric effect $\Delta_{\rm max}$ and for
any pair of sources.
Here, $\theta_i$ are the angles between the source $i$
($i={\mathrm A}, {\mathrm B}$) and the direction of GW propagation. This simply means that
the differential effects cannot exceed the sum of the absolute effects
for the two involved sources. Equation~\eqref{estimate2} implies
that
\begin{equation}
  \label{estimate-2DeltaMax}
  |\delta\psi_\mathrm{AB}|\le 2\Delta_{\rm max}~,
\end{equation}
which gives the upper limit for the GW-induced changes in angular
separations of arbitrary pairs of sources.

While Eq.~\eqref{estimate2} is correct for any pair of sources, one
can derive a better estimate for source pairs with a maximal angular
distance below a certain limit, $\varepsilon$, so that
$\psi_\mathrm{AB}\le\varepsilon$. 
Here, we consider $\varepsilon$ to be
sufficiently small, so that the effect of order $\varepsilon^2$
can be neglected.
Interestingly, one can demonstrate that for arbitrarily
small $\varepsilon$ the estimate given by Eq.~\eqref{estimate2} is almost reachable
(this happens when the center of the field of view coincides with the
direction of the GW propagation, or the opposite
direction). Nevertheless, Eq.~\eqref{estimate2} is, in most cases, overly coarse
and a considerably better estimate can be derived.

The GW-disturbed positions read
$(\alpha_i+\delta\alpha_i,\delta_i+\delta\delta_i)$, $i =
\mathrm{A},\mathrm{B}$, where $\delta\alpha_i$ and $\delta\delta_i$ are
given by Eq.~(A.11) of \cite{2025A&A...695A.172G}.  Using the standard
formula for the angular distance between two points and applying it to both the disturbed and undisturbed source
positions, we arrive at a first-order approximation in $\Delta_{\max}$
(implying also first order in $\delta\psi_\mathrm{AB}$):
\begin{eqnarray}
  \label{deltapsi-epsilon-taylor}
  \delta\psi_\mathrm{AB}=-(\sin\psi_\mathrm{AB})^{-1}\,\bigl(&&
  \left(\cos\delta_\mathrm{A}\sin\delta_\mathrm{B}-\sin\delta_\mathrm{A}\cos\delta_\mathrm{B}\cos(\alpha_\mathrm{B}-\alpha_\mathrm{A})\right)\,\delta\delta_\mathrm{A}
    \nonumber\\
    &+&
  \left(\sin\delta_\mathrm{A}\cos\delta_\mathrm{B}-\cos\delta_\mathrm{A}\sin\delta_\mathrm{B}\cos(\alpha_\mathrm{B}-\alpha_\mathrm{A})\right)\,\delta\delta_\mathrm{B}  
    \nonumber\\
    &-&
     \cos\delta_\mathrm{A}\cos\delta_\mathrm{B}\sin(\alpha_\mathrm{B}-\alpha_\mathrm{A})\,(\delta\alpha_\mathrm{B}-\delta\alpha_\mathrm{A})  
  \bigr)\,.
\end{eqnarray}
Then, considering the variations to first order in $\varepsilon$, one finds the upper estimate
\begin{eqnarray}
  \label{deltapsi-epsilon-estimate}
|\delta\psi_\mathrm{AB}|\le \varepsilon\,\Delta_{\rm max}\,
  \left[1+{\left(1-e^2\,\sin^2(2\overline{\alpha}_r-\phi)\right)}^{1/2}\,(1-\cos\theta_r)\right]+{\cal O}(\varepsilon^2)\,.
\end{eqnarray}
Note that, since one has $\sin\psi_\mathrm{AB} =
\mathcal{O}(\varepsilon)$ in the denominator of
Eq.~\eqref{deltapsi-epsilon-taylor}, the terms of second order in
$\varepsilon$ have been considered in its numerator.  Here,
$\phi$ is the position angle of the elliptic
astrometric signal caused by the GW defined by Eq.~(A.18) of
\cite{2025A&A...695A.172G}.  
Furthermore, $\theta_r$ is
the angular distance between one of the sources, A or B, used as the
reference point with coordinates $(\alpha_r, \delta_r)$ and the
propagation direction of the GW
\begin{eqnarray}
  \label{cos-theta}
\cos\theta_r=\sin\delta_r\sin\delta_{\rm GW}+\cos\delta_r\cos\delta_{\rm GW}\cos\left(\alpha_r-\alpha_{\rm GW}\right)
\end{eqnarray}
and $\overline{\alpha}_r$ is the right ascension of
the reference point in the coordinate system in which the GW propagates toward the north pole:
\begin{equation}
  \begin{pmatrix}
    \cos\overline{\alpha}_r\cos\overline{\delta}_r\\
    \sin\overline{\alpha}_r\cos\overline{\delta}_r\\
    \sin\overline{\delta}_r\\
  \end{pmatrix}
  ={\vec P}^{\rm T}
  \begin{pmatrix}
    \cos\alpha_r\cos\delta_r\\
    \sin\alpha_r\cos\delta_r\\
    \sin\delta_r\\
  \end{pmatrix}\,,
\end{equation}
where ${\vec P}^{\rm T}$ is the transposed matrix $\vec{P}$ given by
Eq.~(A.10) of \cite{2025A&A...695A.172G}. Since
Eq.~\eqref{deltapsi-epsilon-estimate} neglects terms quadratic in
$\varepsilon$, the reference point $(\alpha_r,\delta_r)$ can be
formally taken as the coordinates of the first source
$(\alpha_\mathrm{A},\delta_\mathrm{A})$.

The upper estimate~(\ref{deltapsi-epsilon-estimate}) cannot be
improved for $e=0$, and $e=1$, or when $\sin(2\overline{\alpha}-\phi)=0$ for any value of $e$:
for a given GW with such parameters and for a given $(\alpha_\mathrm{A},\delta_\mathrm{A})$ and $\varepsilon$, one can
find a moment of time and a position
$(\alpha_\mathrm{B},\delta_\mathrm{B})$ for which $|\delta\psi_\mathrm{AB}|$ is exactly given by the right-hand side of Eq.~\eqref{deltapsi-epsilon-estimate}.
Expectedly, one can also demonstrate that the
estimate given by Eq.~\eqref{deltapsi-epsilon-estimate} is reached for pairs with
maximal allowed angular distance $\psi_{\mathrm{AB}}=\varepsilon$.

For $0<e<1$ and $\sin(2\overline{\alpha}-\phi) \neq 0$,
Eq.~\eqref{deltapsi-epsilon-estimate} cannot be exactly attained and can, in principle, be improved.
For the {\sl particular case} of $\sin^2(2\overline{\alpha}-\phi)=1$
one can derive the following reachable estimate:
\begin{eqnarray}
  \label{delta-psi-for-sin^2=1}
  |\delta\psi_\mathrm{AB}|_{\sin^2=1}\le \varepsilon\,\Delta_{\rm max}\,\left[
    \displaystyle{
      \begin{matrix}
    {\left(1+(e^{-2}-1)\,(1-\cos\theta_r)^2\right)}^{1/2},&\qquad (e^{-2}-1)\,(1-\cos\theta_r)\le1\,,\\[5pt]
    \sqrt{1-e^2}\,(2-\cos\theta_r),&\qquad {\rm otherwise}\,.
  \end{matrix}
  }
  \right.
\end{eqnarray}
We note that Eq.~\eqref{delta-psi-for-sin^2=1} gives exactly the same
estimate for $|\delta\psi_\mathrm{AB}|$ as
Eq.~\eqref{deltapsi-epsilon-estimate} for $e=0$ and $e=1$ (and
$\sin^2(2\overline{\alpha}-\phi)=1$).  A combined analytical and
numerical investigation shows that
Eq.~\eqref{deltapsi-epsilon-estimate} overestimates the real maximal
value of $|\delta\psi_\mathrm{AB}|$ by at most a factor of 1.4. Even
if a better general estimate could be given as a complicated function
of $e$, $\cos\theta_r$ as well as the sine and cosine of
$2\overline{\alpha}_r-\phi$, we prefer to use
Eq.~\eqref{deltapsi-epsilon-estimate} because of its simplicity.

We note that one can further simplify Eq.~\eqref{deltapsi-epsilon-estimate} as
\begin{equation}
  \label{estimate-2-costheta}
|\delta\psi_\mathrm{AB}|\le \varepsilon\,\Delta_{\rm max}\,(2-\cos\theta_\mathrm{r})
\end{equation}
to make it valid for any eccentricity $e$ (this is exactly Eq.~(\ref{deltapsi-epsilon-estimate}) for $e=0$) and,
finally, as
\begin{equation}
  \label{estimate-3DeltaMaxEpsilon}
|\delta\psi_\mathrm{AB}|\le 3 \varepsilon\,\Delta_{\rm max}\,.
\end{equation}
This gives the upper estimate of the GW-induced variation of the
angular distance for any $e$ and for any position on the sky. Since
$\varepsilon$ is considered to be sufficiently small (in principle,
$\varepsilon\ll1$) this latter estimate does not contradict
Eq.~\eqref{estimate-2DeltaMax}. Both Eq.~\eqref{estimate-2-costheta}
and Eq.~\eqref{estimate-3DeltaMaxEpsilon} are reachable in the
respective parameter space.

Finally, we point out two interesting aspects of
$|\delta\psi_\mathrm{AB}|$. First, since the absolute astrometric
effect of a GW is proportional to $\sin\theta$ (e.g. it is maximal at
the angular distance of $\theta=\pi/2$ from the GW propagation
direction) one could, naively, expect similar dependence of the
differential effect $\delta\psi_\mathrm{AB}$ also for source pairs
with small angular distance $\varepsilon$. The estimate given
by Eq.~\eqref{estimate2} seems to support this expectation. However,
we see from Eq.~\eqref{estimate-2-costheta} that the maximal value of
$|\delta\psi_\mathrm{AB}|$ is proportional to $2-\cos\theta$. This
means that the maximal differential effect is, in fact, minimal in the
direction of the GW propagation, where it reaches
$\varepsilon\,\Delta_{\rm max}$. It then continuously increases up to
$3\varepsilon\,\Delta_{\rm max}$ towards the direction of the GW
source ($\theta=\pi$). This is also illustrated by
Figure~\ref{fig__non_histogram_theta}.

Another remarkable aspect of $|\delta\psi_\mathrm{AB}|$ is the
existence of a flower-like pattern with four ``petals'' for GWs with $e > 0$, as one can
see in Figure~\ref{fig__skymaps_and_non_histograms}.
This pattern is related to the term in Eq.~\eqref{deltapsi-epsilon-estimate}
that depends on $\overline{\alpha}_\mathrm{A}$.
For four values of $\overline{\alpha}_\mathrm{A}$ for
which $\sin(2\overline{\alpha}_\mathrm{A}-\phi) = 0$ the maximal
differential effect is given by Eq.~\eqref{estimate-2-costheta},
while for the other values of the sine the maxima of
$|\delta\psi_\mathrm{AB}|$ for $\theta = \pi$ become shallower and reach
their minimal values $\varepsilon\,\Delta_{\rm max}\,(1+2\sqrt{1-e^2})$ for
$\sin(2\overline{\alpha}_\mathrm{A}-\phi) = \pm1$.
We note that for $e = 1$ and $\sin(2\overline{\alpha}_\mathrm{A}-\phi) = \pm1$ one gets
$|\delta\psi_\mathrm{AB}|\le \varepsilon\,\Delta_{\rm max}$ independently
of $\theta$.

This four-petalled pattern in the differential astrometric effect has
an important consequence for the choice of the reference point in
Eq.~\eqref{deltapsi-epsilon-estimate}. The linear approximation in
$\varepsilon$ used in Eq.~\eqref{deltapsi-epsilon-estimate} is
sufficient when the effect is not changing much on the scale of
$\epsilon$ across the sky. However, if the sources are located close to
the GW source, the four-petalled pattern shown in the two lower plots
of Fig.~\ref{fig__skymaps_and_non_histograms} changes very quickly,
and the factor
${\left(1-e^2\,\sin^2(2\overline{\alpha}-\phi)\right)}^{1/2}$ that is
computed for the reference point $(\alpha_r,\delta_r)$ for $r =
\mathrm{A}$ in Eq.~\eqref{deltapsi-epsilon-estimate} can be
significantly different when computed for the other point with $r =
\mathrm{B}$.  Our numerical studies show that in some extreme cases,
when the distances of both sources from the GW source are comparable to
$\varepsilon$, and if $(\alpha_\mathrm{A},\delta_\mathrm{A})$ is
chosen for $(\alpha_r,\delta_r)$ as discussed above, the estimate from
Eq.~\eqref{deltapsi-epsilon-estimate} can give a slightly lower value
than the actual $|\delta\psi_\mathrm{AB}|$.  To cover these cases, the
reference point $(\alpha_r,\delta_r)$ should be chosen to be either
$(\alpha_\mathrm{A},\delta_\mathrm{A})$ or
$(\alpha_\mathrm{B},\delta_\mathrm{B})$, whichever gives the smaller
value of ${\left(1-e^2\,\sin^2(2\overline{\alpha}-\phi)\right)}^{1/2}$
in Eq.~\eqref{deltapsi-epsilon-estimate}.

Eq.~\eqref{deltapsi-epsilon-estimate} gives a reasonable upper estimate
for pairs of sources with maximal angular distance $\varepsilon$. For
all possible pairs of sources within a round field of view with
angular diameter $\varrho$, Eq.~(\ref{deltapsi-epsilon-estimate}) remains valid for
$\varepsilon=\varrho$. Similarly, for all pairs of sources within a
square field of view of angular size $\varrho\times\varrho$,
Eq.~\eqref{deltapsi-epsilon-estimate} is valid for
$\varepsilon=\sqrt{2}\,\varrho$.

Overall, we conclude that the differential effect $\delta\psi_\mathrm{AB}$
in the angular distance between two sources remains, as expected, of the
same order of magnitude as the absolute effect discussed e.g. by
\cite{Article_Book_Flanagan,2025A&A...695A.172G}:
$|\delta\psi_\mathrm{AB}|\le 2\Delta_{\rm max}$ for arbitrary pairs of stars.
However, for limited FoVs the differential effect is limited to $|\delta\psi_\mathrm{AB}|\le 3\varepsilon\,\Delta_{\rm max}$ for any
pairs of sources at the angular distance of $\varepsilon$ or lower.

The authors of \cite{MT_Crosta} and \cite{2025NatSR..1532908S}
predicted an increasingly large astrometric effect from GWs at small
separations. This is not confirmed by our analysis. The
technical reasons for the flaw in their work are described in the Appendix (see also
\cite{2025arXiv250718593V}, where this flaw is discussed as well).
 
\section{Numerical Simulations}
\label{section-numerical}

In order to verify and visualize the findings from
Section~\ref{sec__max_change} we conducted a series of numerical
simulations in which we explicitly compute the differential astrometric
GW effect $\delta\psi_\mathrm{AB}$ between two stars~${\mathrm A}$~and~${\mathrm B}$ for specific angular distances.

As the theoretical discussion above suggests, and our numerical simulations
confirm, the particular values of the maximum
possible angular separation, $\varepsilon$, of source pairs, as well as
the maximal GW amplitude, $\Delta_{\rm max}$, can be chosen
arbitrarily as long as they remain sufficiently small: the effects are
always proportional to $\varepsilon\,\Delta_{\rm max}$. In
particular, this remains true unless we consider very large
$\varepsilon$ of many degrees, for which the second-order effects neglected
in Section~\ref{sec__max_change} become numerically important. To be
realistic for small-field astrometry, for all simulations, we used
$\varepsilon = 0.1^\circ$ and $\Delta_{\rm max} =
10\,\mathrm{mas}\approx 4.8 \times 10^{-8}$. The values are small
enough that second-order effects can be neglected, and at the same
time, the magnitude of the effect is large enough so that numerical
noise is not an issue.
It is clear that in reality $\Delta_{\rm max}$ will certainly be much smaller, most likely in the region of nano-arcseconds and below.

All results presented below are given as a normalized angular change,
$\mathcal{F} = \delta\psi_\mathrm{AB} /
(\varepsilon\,\Delta_\mathrm{max})$. In the linear approximation,
which we consider, the value of $\mathcal{F}$ is independent of
$\varepsilon$ and $\Delta_\mathrm{max}$. Conversely, the corresponding
magnitude of the angular change can be restored as
$\delta\psi_\mathrm{AB}=\mathcal{F}\,
\varepsilon\,\Delta_\mathrm{max}$ for a particular FoV size
$\varepsilon$ and a GW with a maximal astrometric effect of $\Delta_\mathrm{max}$.

\subsection{Basic statistics of a typical differential signal}
\label{sec__basic_statistics}

To get a first coarse overview, we computed some statistics of $\mathcal{F}$
as one might expect from a random selection of stars and GW parameters.
This, in a way, also reflects our lack of a priori knowledge about specific GW emitters.
To this end, we simulated $10^9$ sets of randomly selected GW parameters.
After random selection, the strain parameters were always scaled in such a way that $\Delta_{\rm max}$ remained constant.
For each GW, we selected 100 random source pairs across the celestial sphere;
each source pair had a randomly selected angular separation of $0 <
\psi_\mathrm{AB} \leq \varepsilon$ and a random orientation. For each
of the pairs, we computed the change in angular distance due to a GW
at a random time. This gives $10^8$ overall
samples of $\mathcal{F}$, the basic statistics of which are given in
Table~\ref{tab__stats_for_random}. 
First, we see that our simulations confirm that $\left|\mathcal{F}\right|\le3$
as suggested by Eq.~\eqref{estimate-3DeltaMaxEpsilon}.
We also see that a typical value of $\left|\mathcal{F}\right|$ is 0.18, which means
that a typical value of $|\delta\psi_\mathrm{AB}|$ in the random small-field astrometric observations
is about $0.18\,\varepsilon\,\Delta_{\rm max}$. For a FoV with, e.g.,
$\varepsilon=0.1^\circ$, this implies that a typical sensitivity of only 0.03\% 
of the already minuscule $\Delta_{\rm max}$ would be necessary to
detect such a typical signal.

\begin{table}[htb]
\begin{minipage}{0.975\linewidth}
		\caption{Statistics of normalized absolute angular changes $\mathcal{F}$
			from the simulated data using random star pairs and random GW parameters.
		  The values are, from top to bottom, the mean, the standard deviation, and
		  a series of quantiles: the minimum, the 0.1, 0.5, and 0.9 quantiles
		  and the maximum of the absolute value of
		  $\mathcal{F}$.\label{tab__stats_for_random}}
\end{minipage}
~~\\[1.5ex]
\begin{minipage}{0.4\linewidth}
\begin{ruledtabular}
	\begin{tabular}{lr}
		\textrm{Parameter} & \textrm{Value} \\
		\colrule
		\noalign{\vskip 2pt}
		$\mathrm{mean}\left(|\mathcal{F}|\right)$ & 0.30 \\
		$\mathrm{std}\left(\mathcal{F}\right)$	& 0.46 \\[3pt]
                $\min{\left(|\mathcal{F}|\right)}$ & 0.00 \\
		$Q_{0.1}\left(|\mathcal{F}|\right)$  & 0.02 \\
                $\mathrm{median}\left(|\mathcal{F}|\right)$ & 0.18 \\
		$Q_{0.9}\left(|\mathcal{F}|\right)$  & 0.76 \\
                $\max{\left(|\mathcal{F}|\right)}$ & 2.98 \\
	\end{tabular}
\end{ruledtabular}
\end{minipage}
\end{table}

\subsection{Spatial distribution of the differential astrometric GW signal}

Next, we consider the spatial distribution of the variations in
angular distances $|\delta\psi_\mathrm{AB}|$ for a given GW. The
spatial distribution of the signal is especially relevant for small-field
astrometry, given that only a limited number of sky regions can
typically be observed. If, for instance, a GW emitter candidate is
identified beforehand, observations might be optimally directed towards the area
of the sky with the highest probability of detecting the GW from this
emitter.

We note that the propagation direction of the GW can be chosen
arbitrarily, since any other GW direction is equivalent to a different
orientation of the coordinate system. Since we are looking for the
maximal values of $|\delta\psi_\mathrm{AB}|$ over an extended period of
time, the GW frequency can also be selected arbitrarily, provided that
the tested time interval covers at least one GW period. Thus,
only four strain parameters are important in this study. As shown in
the Appendix A of \cite{2025A&A...695A.172G}, the magnitude of the
astrometric GW effect can alternatively be described by the maximal
astrometric effect $\Delta_{\rm max}$, the eccentricity $e$,
and the position angle $\phi$ of the ellipse representing the
astrometric GW signal.

One can show that a change of the position angle $\phi$ is equivalent to
a rotation of the reference system around the propagation direction $\ve{p}$ of
the GW.

$\Delta_{\rm max}$ can also be fixed to a constant for the
reasons explained above, and it then remains to examine
the distributions for various values of $e$.

To investigate this, we simulated the signals from three GWs with
eccentricities $e$ equal to 0, 0.7, and 1. For each GW, we generated
$10^8$ randomly selected and oriented source pairs, each possessing a
constant (unperturbed) angular separation $\psi_\mathrm{AB} =
\varepsilon$. We deliberately chose $\psi_\mathrm{AB}$ to be the
maximum pair separation, $\varepsilon$, since we are mainly interested
in the maximal angular change observable in a given area of the sky.
Subsequently, for each source pair, we computed the change in angular
separation $\delta\psi_\mathrm{AB}$ induced by the GW at randomly
selected observation times (over a time span considerably larger
than the GW period). These data were then used to visualize the sky
distribution of the maximal value of $|\delta\psi_\mathrm{AB}|$, and
to validate the dependence of this quantity on $\theta_r$ and
$\overline{\alpha}_r$ as in Eqs.~\eqref{deltapsi-epsilon-estimate} and
\eqref{estimate-2-costheta}.

\begin{figure}[p!]
	\centering
	\begin{tikzpicture}
		\matrix[matrix of nodes,
		nodes={inner sep=0pt, outer sep=0pt, anchor=center},
		column sep=0.2cm, row sep=0.1cm] (m) {
			\node{\begin{minipage}{0.4\textwidth}
					\includegraphics[keepaspectratio,width=1\textwidth,valign=c]{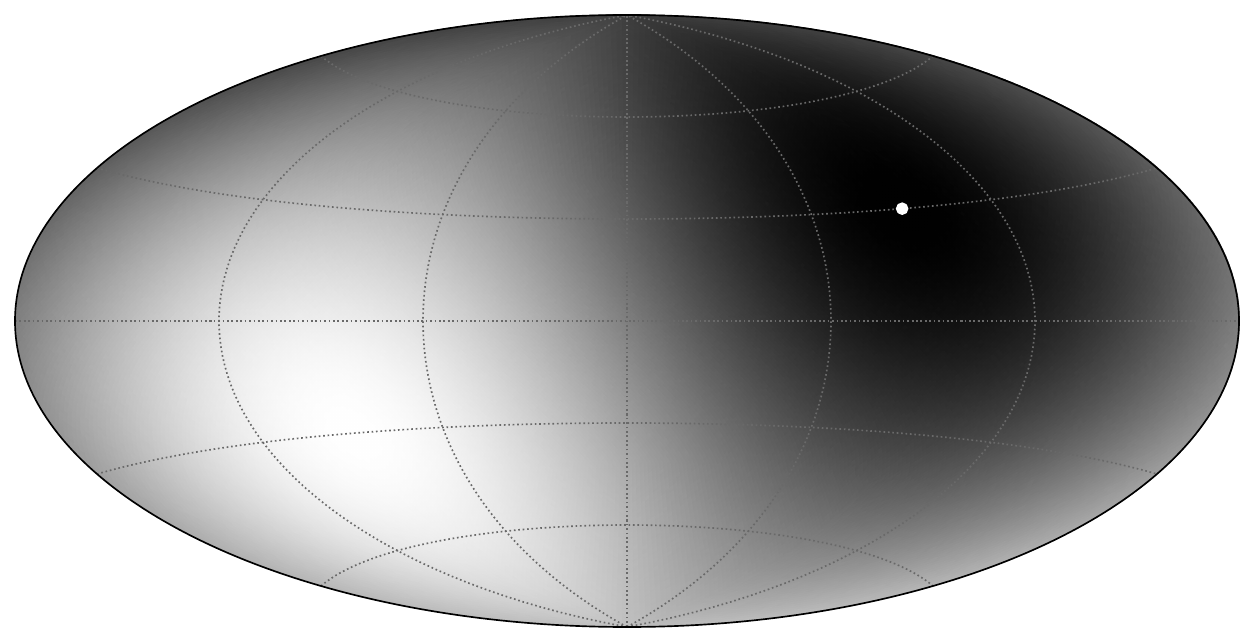}
					\vspace{3.5mm}
				  \end{minipage}}; &
			\node{\includegraphics[keepaspectratio,width=0.4\textwidth,valign=c]{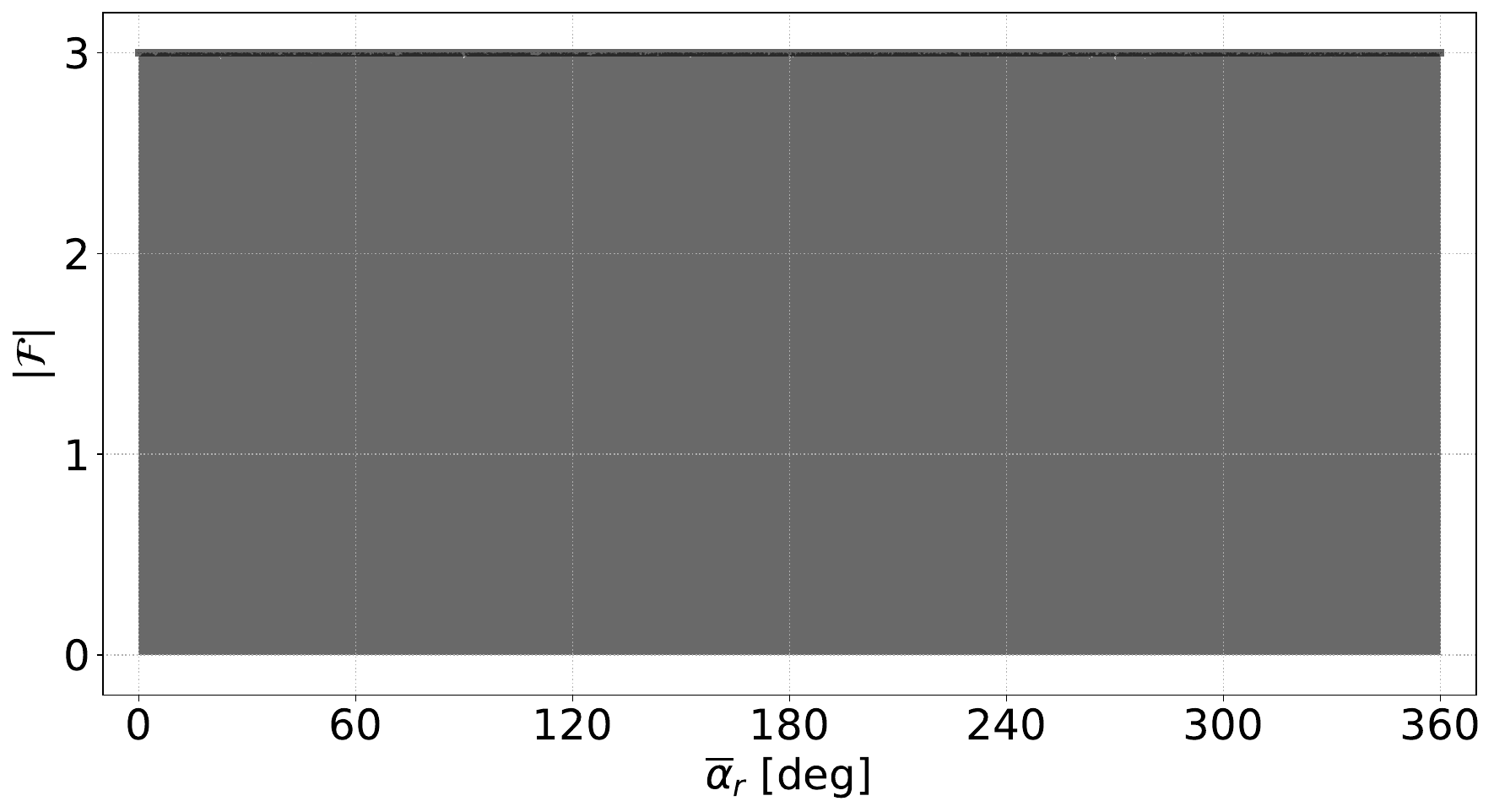}}; \\
			\node{\begin{minipage}{0.4\textwidth}
					\includegraphics[keepaspectratio,width=1\textwidth,valign=c]{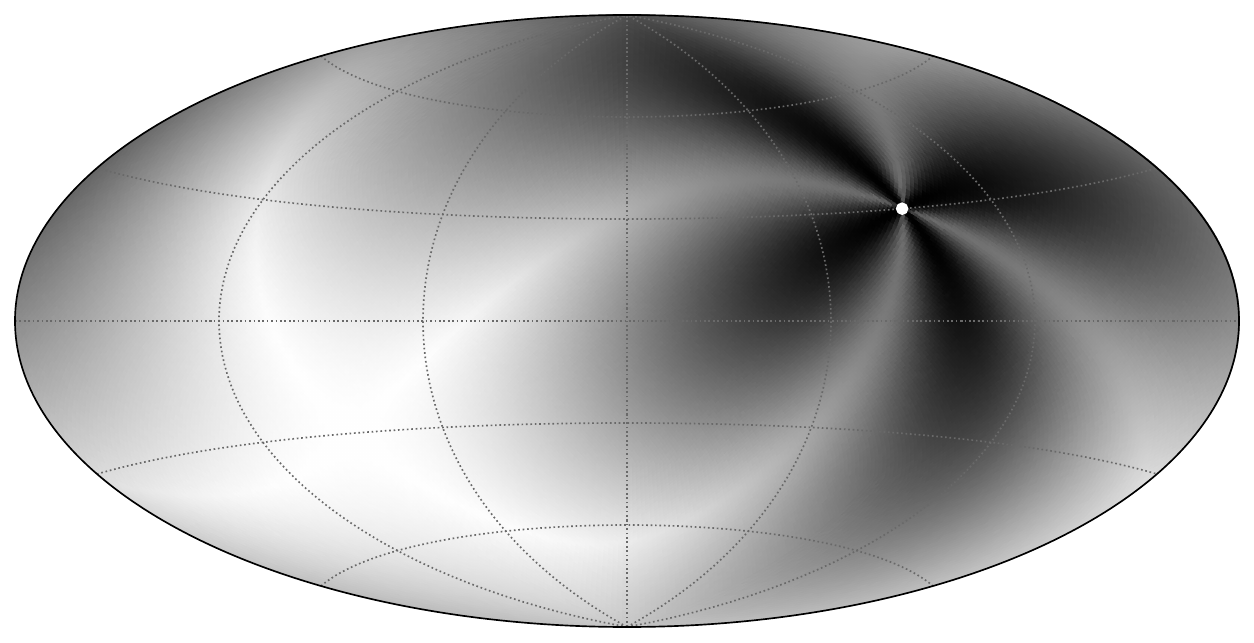}
					\vspace{3.5mm}
				\end{minipage}}; &
			\node{\includegraphics[keepaspectratio,width=0.4\textwidth,valign=c]{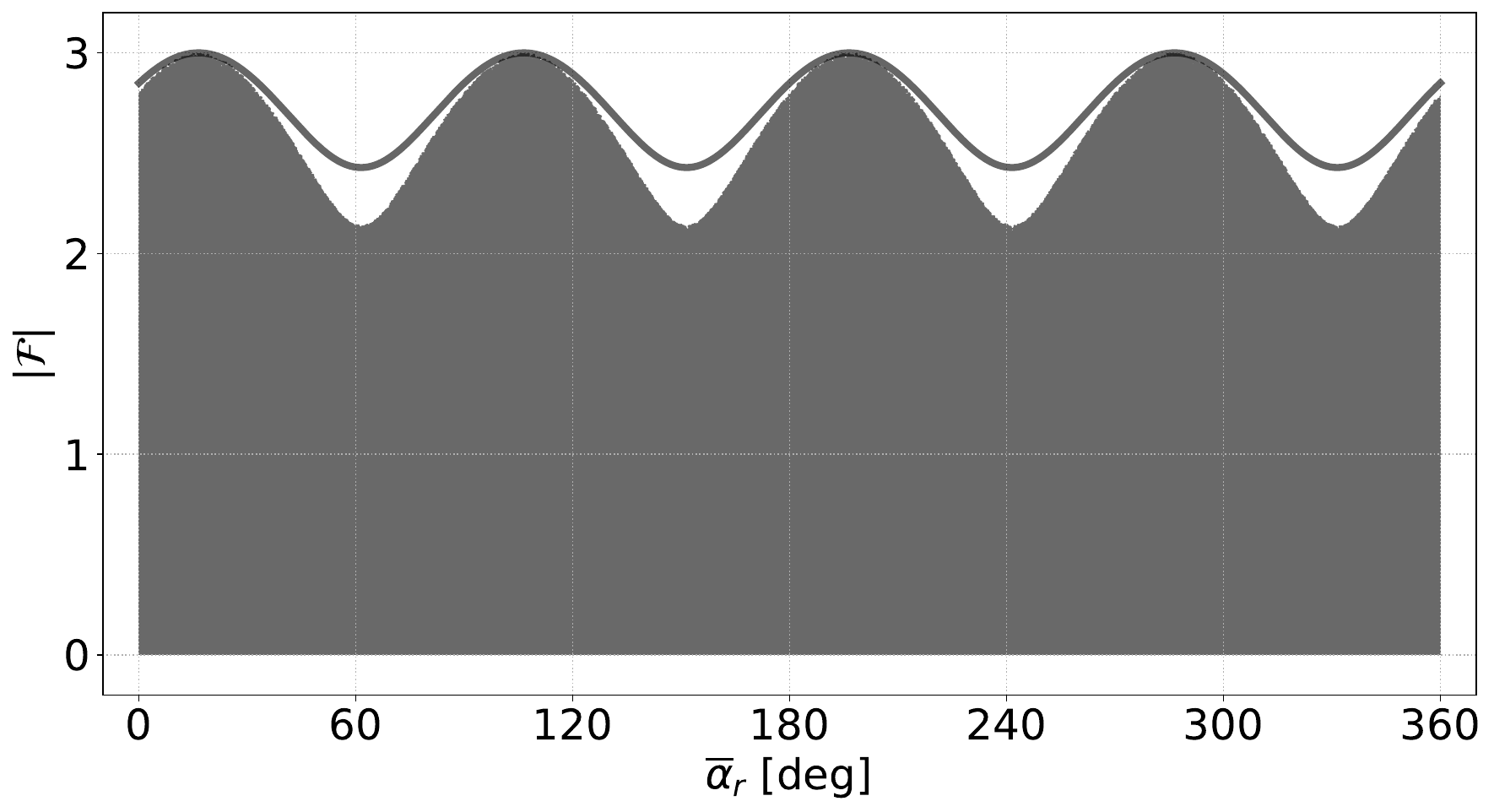}}; \\
			\node{\begin{minipage}{0.4\textwidth}
					\includegraphics[keepaspectratio,width=1\textwidth,valign=c]{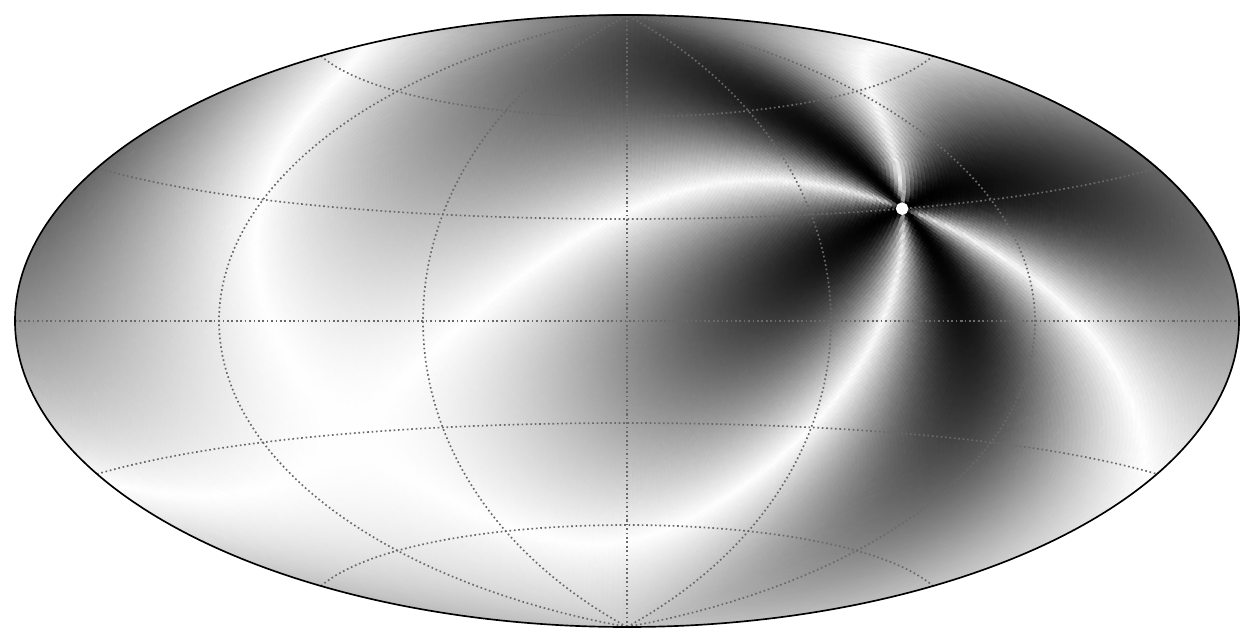}\hfill
					\includegraphics[keepaspectratio,width=1\textwidth,valign=c]{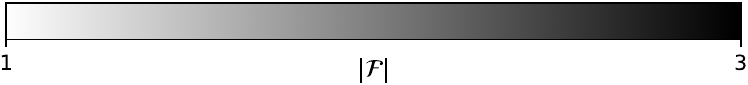}
			\end{minipage}};
			&
			\node{\includegraphics[keepaspectratio,width=0.4\textwidth,valign=c]{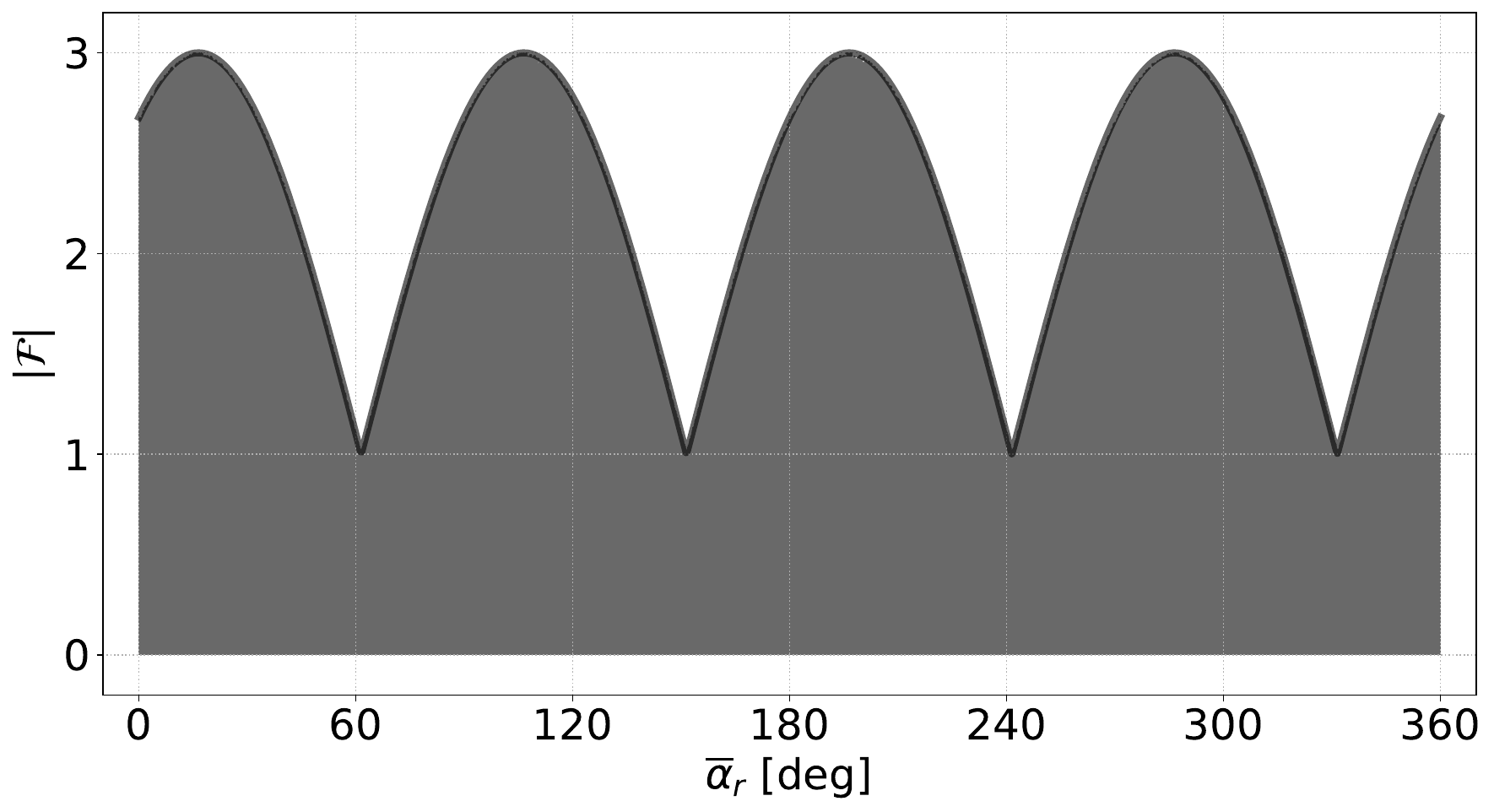}}; \\
	    };
		\end{tikzpicture}
	\caption{Resulting normalized angular changes represented by
          $|\mathcal{F}|$ from numerical simulations using random star
          pairs and three GWs with fixed parameters. The top row
          corresponds to a GW signal with an eccentricity of $e=0$,
          the middle row to $e=0.7$, and the bottom row to $e=1$. The
          left column presents sky maps showing the
          maximum absolute normalized change of angular distance,
          $\max\left(|\mathcal{F}|\right)$, per HEALPix of level
          6. The small white dot in the sky maps marks the position of
          the GW emitter. In the right column, all simulated
          normalized angular changes are displayed as a function of
          $\overline{\alpha}_r$. A black line in these plots marks
          the maximal achievable $|\mathcal{F}|$ according to
          Eq.~\eqref{fig__non_histogram_theta}. The sky maps use the
          Hammer-Aitoff projection in equatorial coordinates, with
          $\alpha = \delta = 0$ at the center, north up, and $\alpha$
          increasing from right to
          left.\label{fig__skymaps_and_non_histograms}}
\end{figure}

\begin{figure}[p!]
	\centering
	\includegraphics[keepaspectratio,width=0.45\textwidth]{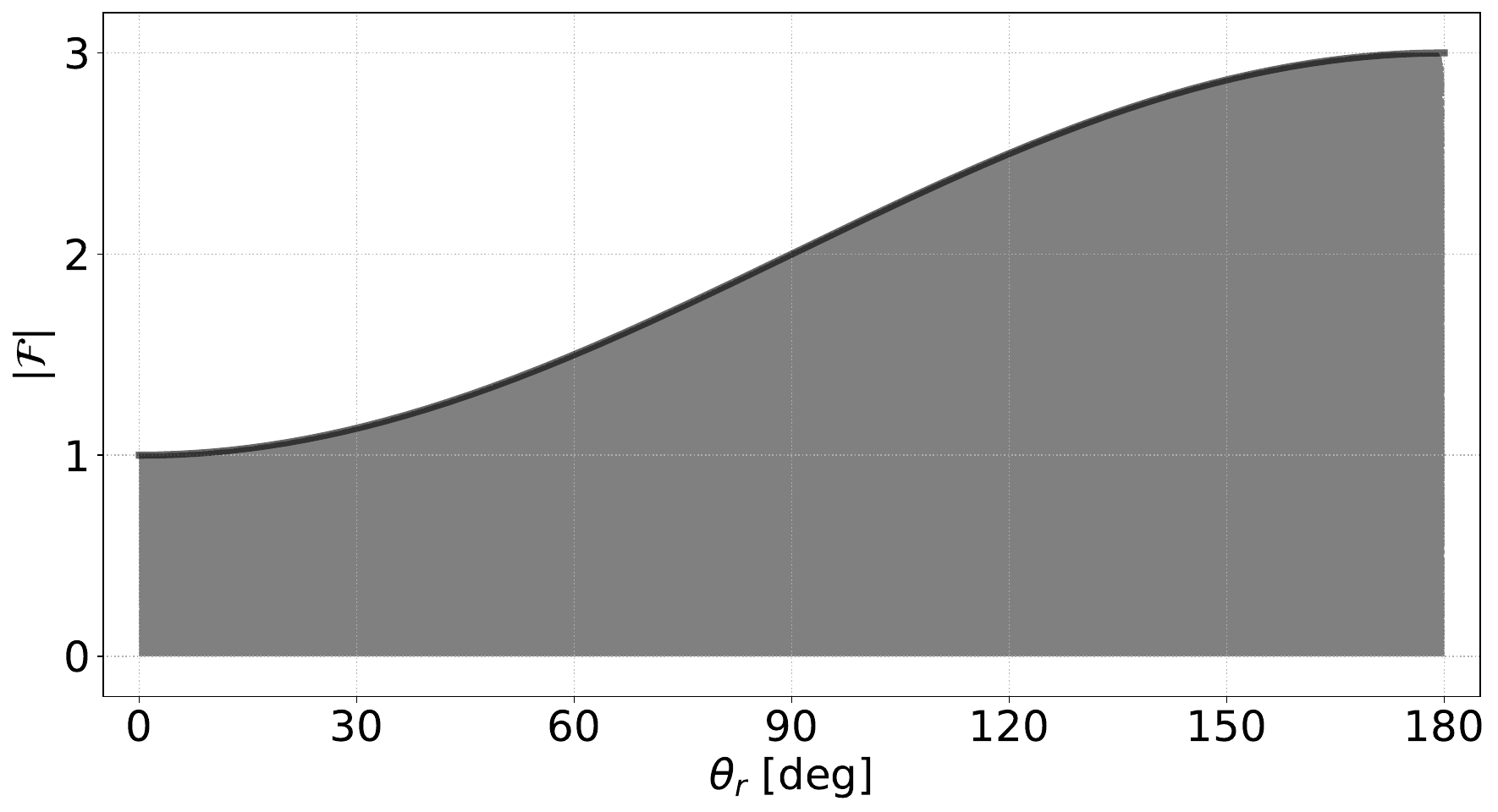}
	\caption{The normalized changes of the angular distance from
          the numerical simulations, $|\mathcal{F}|$, as a function of
          $\theta_r$. This plot is valid for all eccentricities
          $e$. The differences between the three cases shown in
          Fig.~\ref{fig__skymaps_and_non_histograms} are negligible.
          The black line represents the value given by
          Eq.~\ref{estimate-2-costheta}.\label{fig__non_histogram_theta}}
\end{figure}

Figures~\ref{fig__skymaps_and_non_histograms} and
\ref{fig__non_histogram_theta} show the results of the simulations.
The sky maps in Fig.~\ref{fig__skymaps_and_non_histograms} show the
maximal value of $|\mathcal{F}|$ per HEALPix
\cite{2005ApJ...622..759G} on the sky, for all simulated source pairs
falling into the respective pixel.  These maps clearly illustrate the
expected dependence on $\theta_r$ and $\overline{\alpha}_r$, which
form the four-petalled pattern on the sky for $e>0$.  The respective
scatter plots of simulated angular changes over $\overline{\alpha}_r$
in the right column of Fig.~\ref{fig__skymaps_and_non_histograms}
further elucidate the relationship with $e$. It is important to note
that these scatter plots also unambiguously demonstrate that the
actual measured angular change of the source pair can be zero at any
$\overline{\alpha}_r$ because the change in angular separation due to
a GW depends on the orientation of the source pair.  The differential
effect can be zero even at $\overline{\alpha}_r$, where a maximum in
$|\mathcal{F}|$ is reached.

Figure~\ref{fig__non_histogram_theta} illustrates the dependence of
the simulated angular changes on $\theta_r$, the angular distance to the
GW propagation direction. This result agrees
with Eq.~\ref{estimate-2-costheta} for the maximum values.
We note that in this case, too, the actual measured angular change depends on the orientation of the source pair, and it can be zero for any $\theta_r$.

\subsection{Differential astrometric GW signal inside a FoV}

While considering the change in the angular distance of individual pairs
of sources is important, it provides limited insight into the
distribution of the differential astrometric GW signal over the area
of a FoV. To elucidate that distribution, we simulated a GW signal
using a purely $+$-polarized GW, sampled at the time of maximal
magnitude inside a FoV for different points on the sky. This GW
configuration is also representative of the $\times$-polarisation,
although the specific examples would look different, the conclusions
are the same.

\begin{figure}[p!]
	\centering
	\includegraphics[keepaspectratio,height=0.3\textwidth]{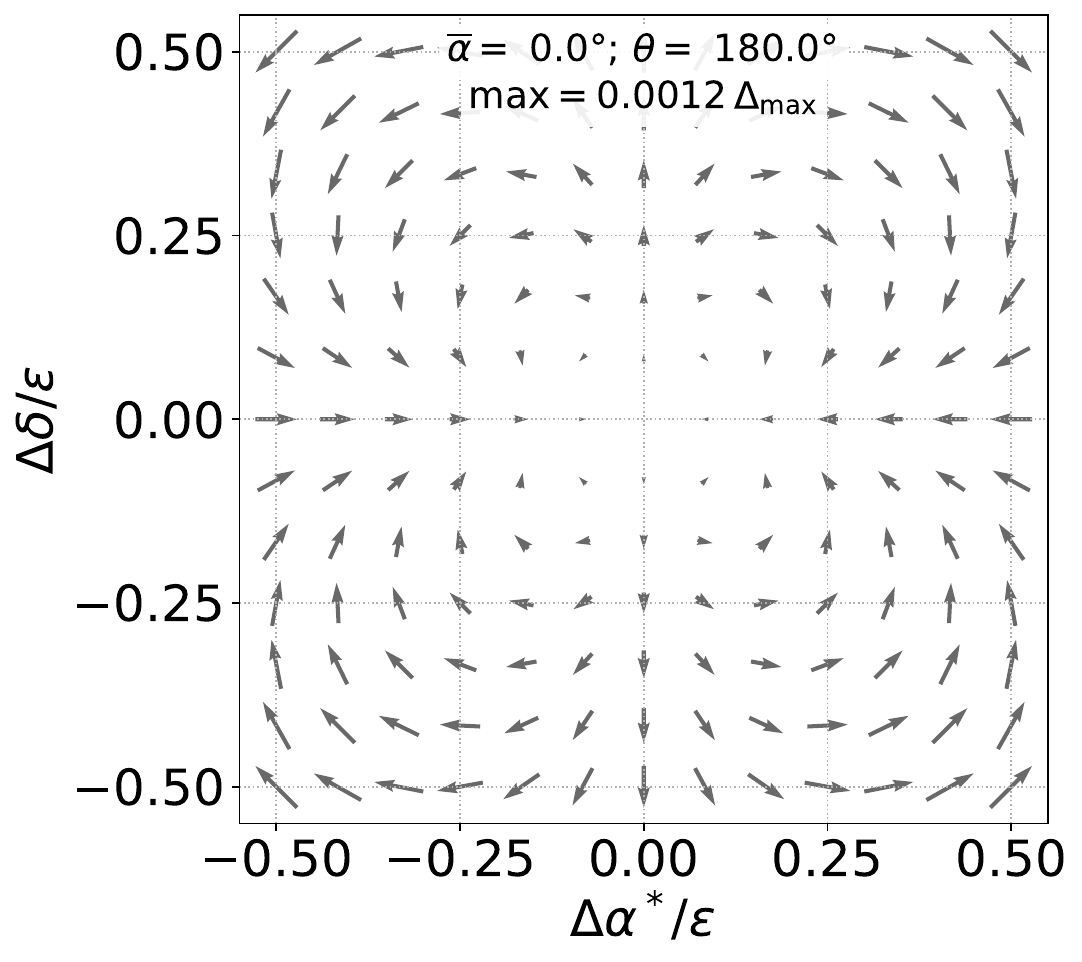}
	\includegraphics[keepaspectratio,height=0.3\textwidth]{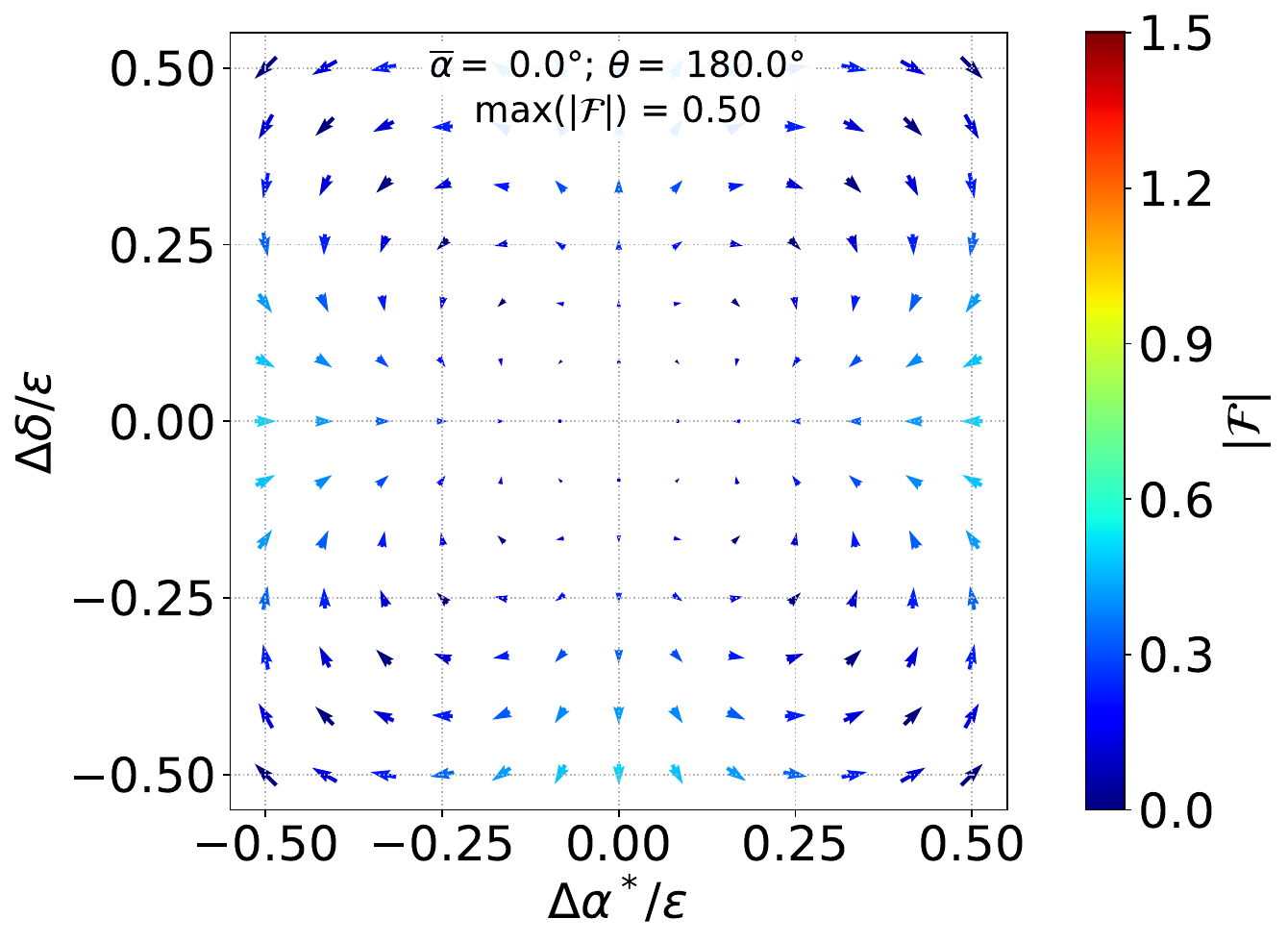}\\
	\includegraphics[keepaspectratio,height=0.3\textwidth]{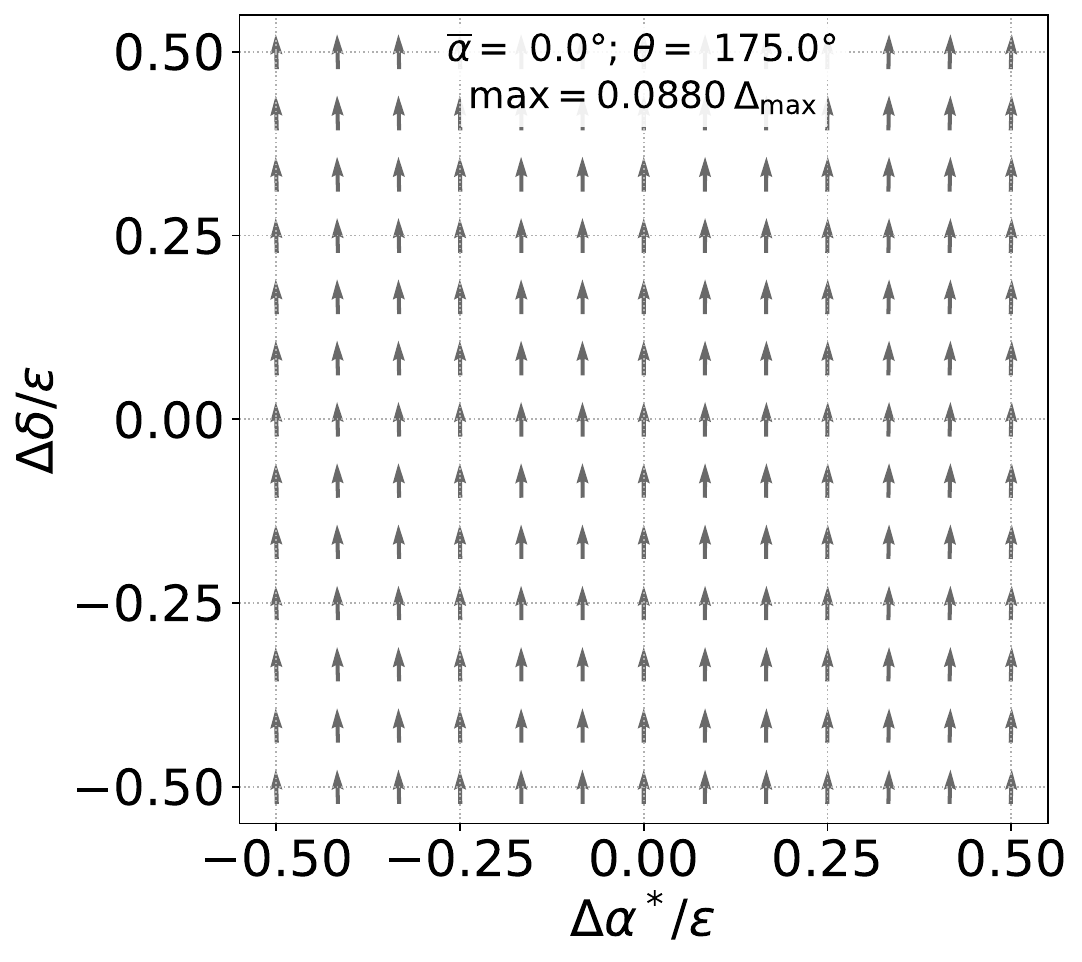}
	\includegraphics[keepaspectratio,height=0.3\textwidth]{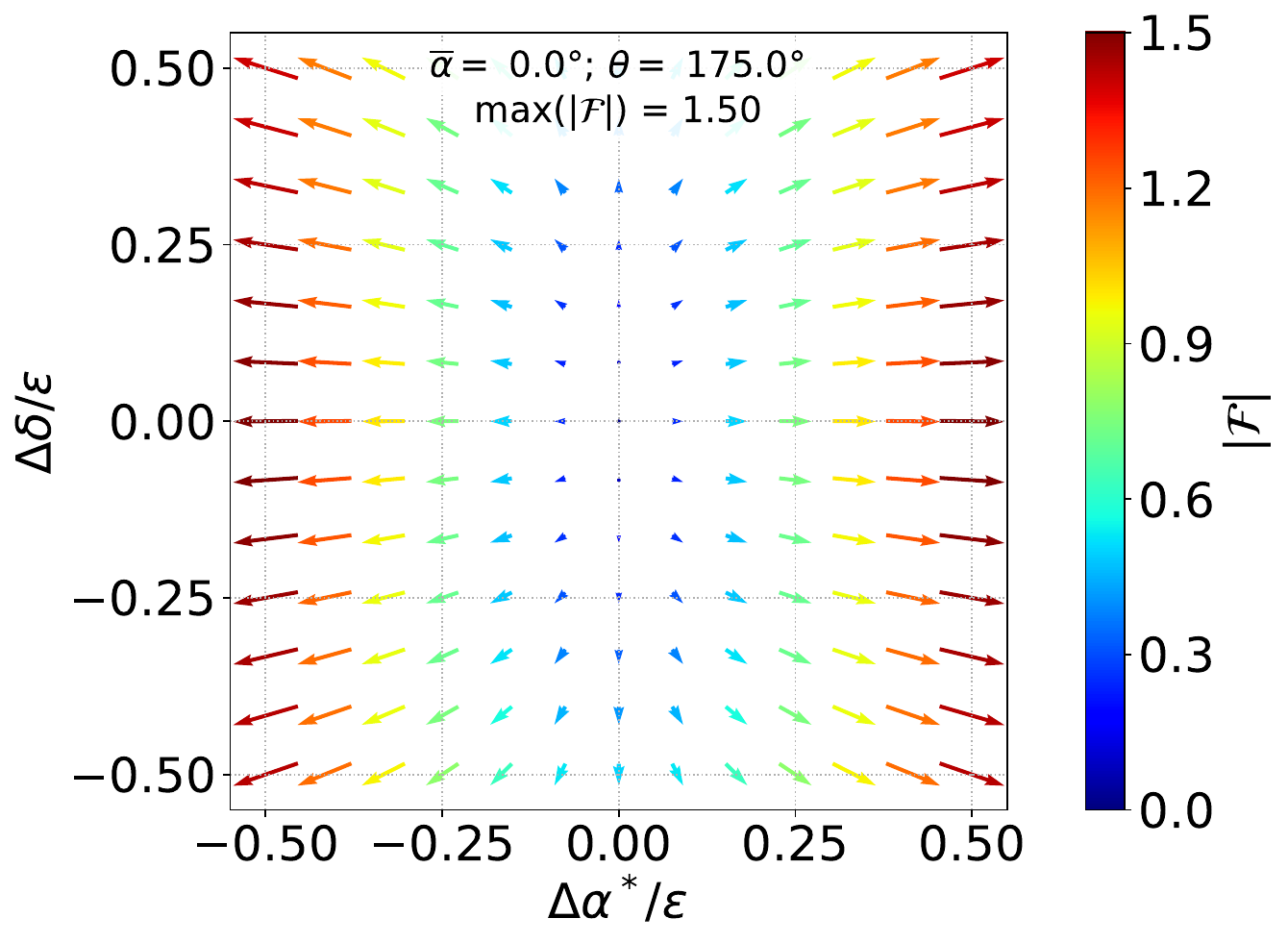}\\
	\includegraphics[keepaspectratio,height=0.3\textwidth]{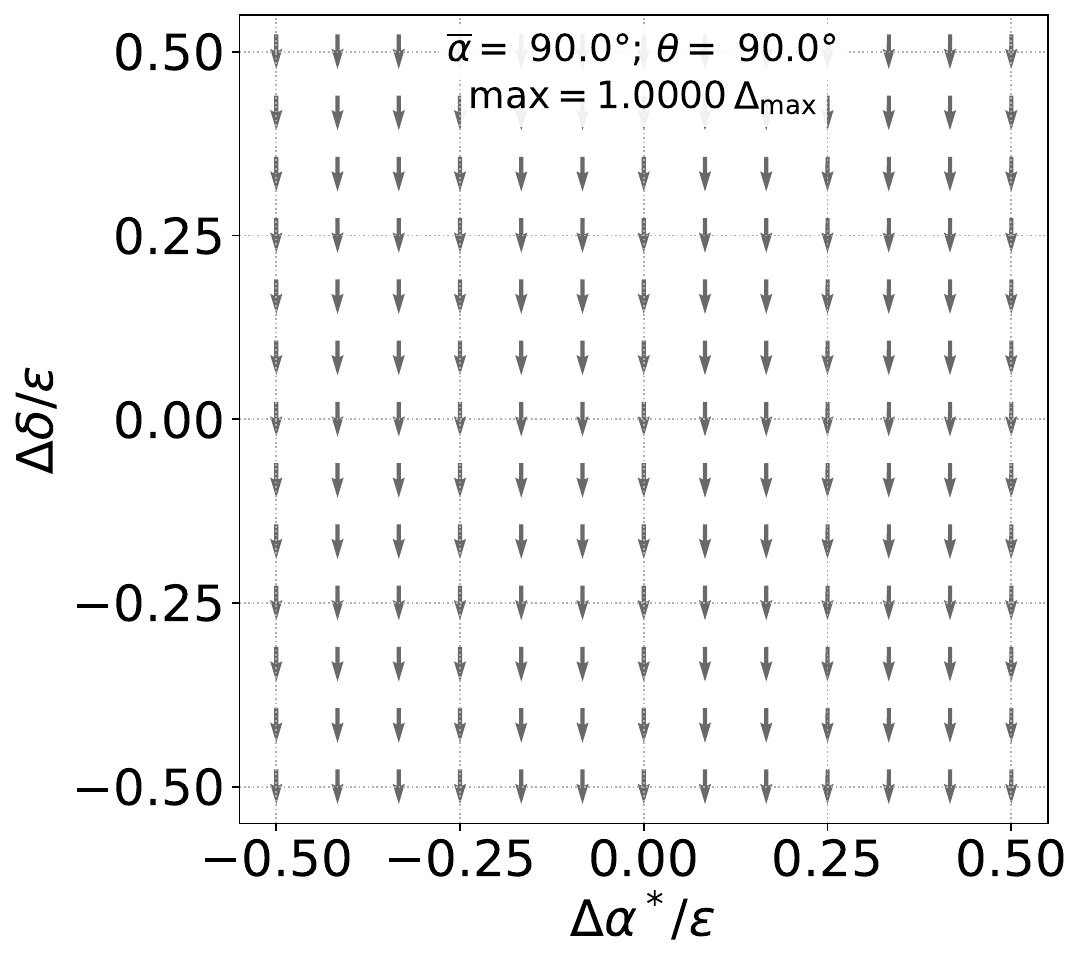}
	\includegraphics[keepaspectratio,height=0.3\textwidth]{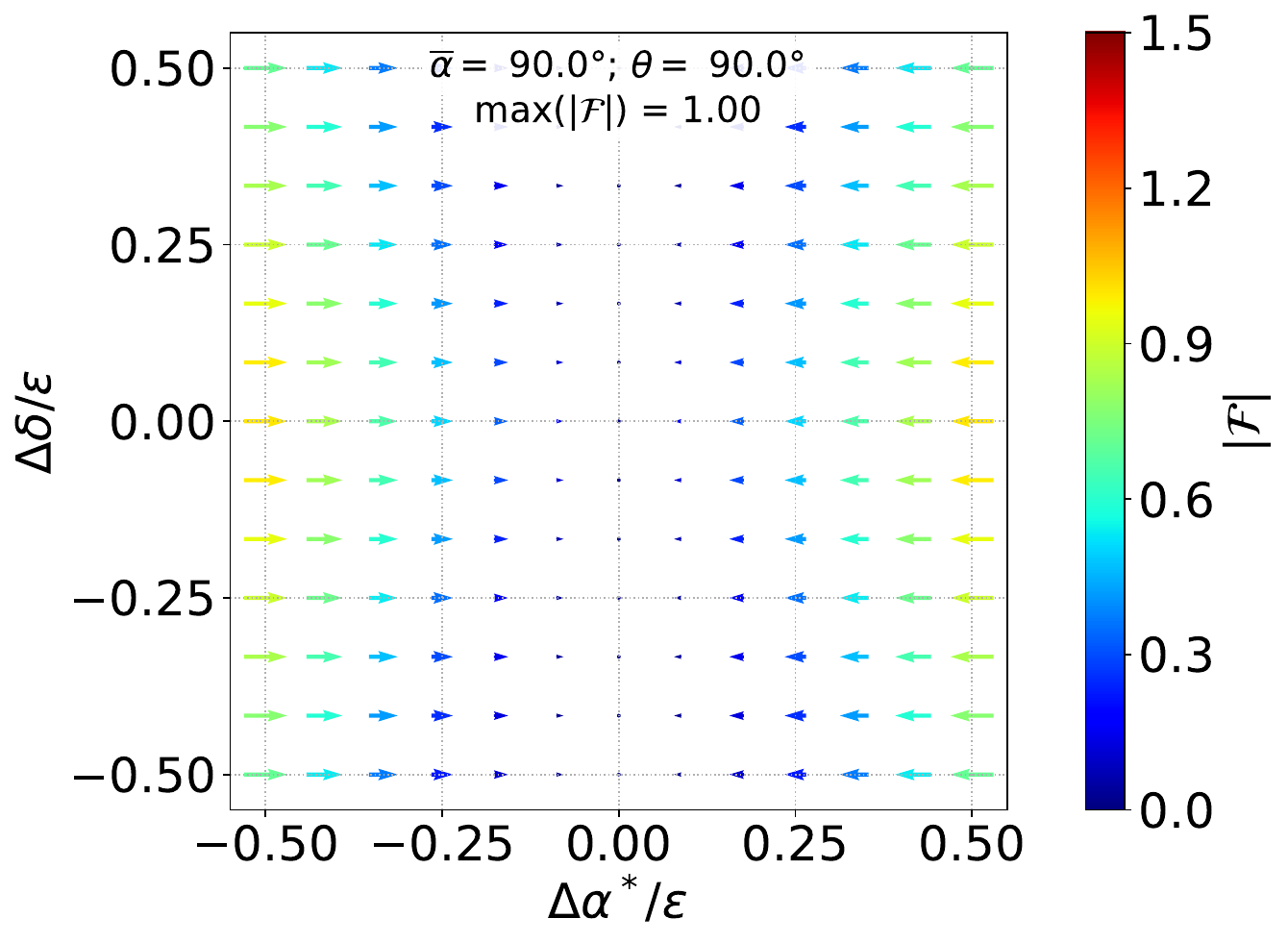}\\
	\includegraphics[keepaspectratio,height=0.3\textwidth]{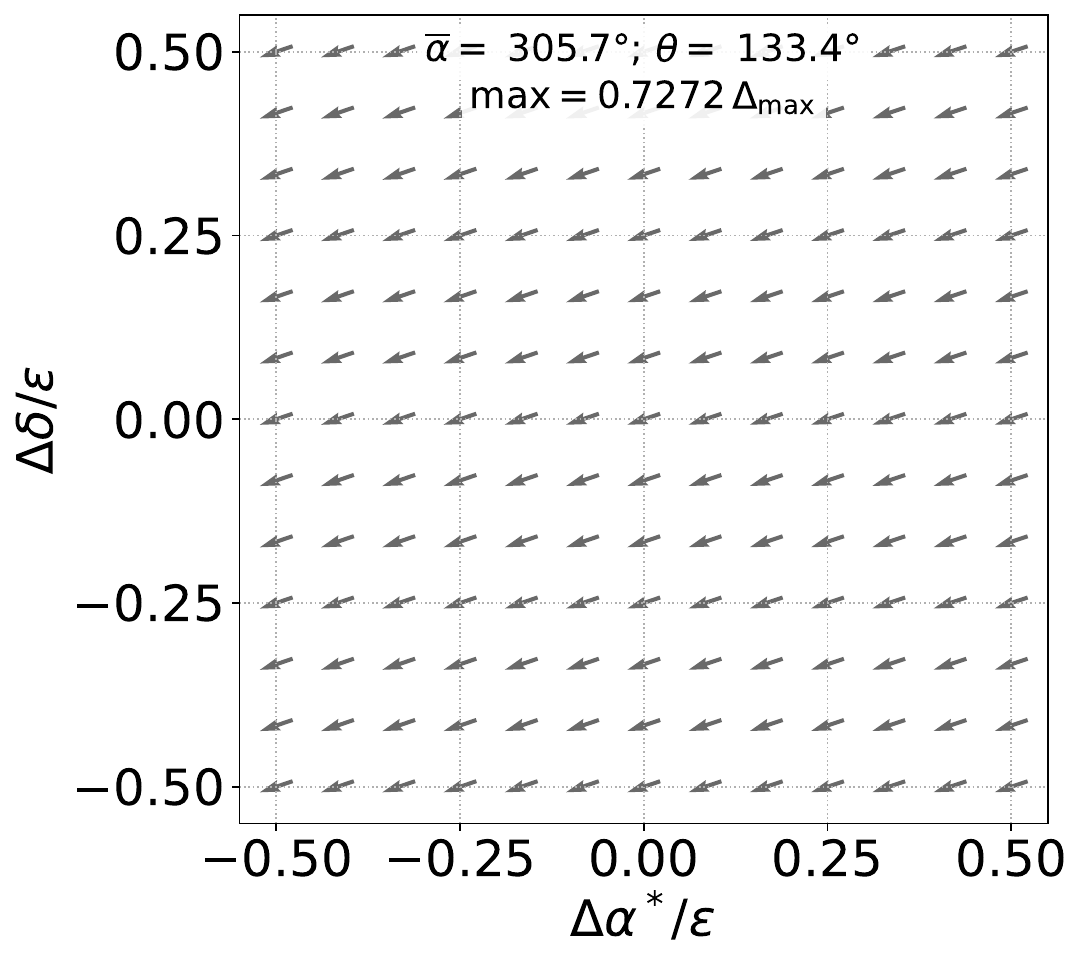}
	\includegraphics[keepaspectratio,height=0.3\textwidth]{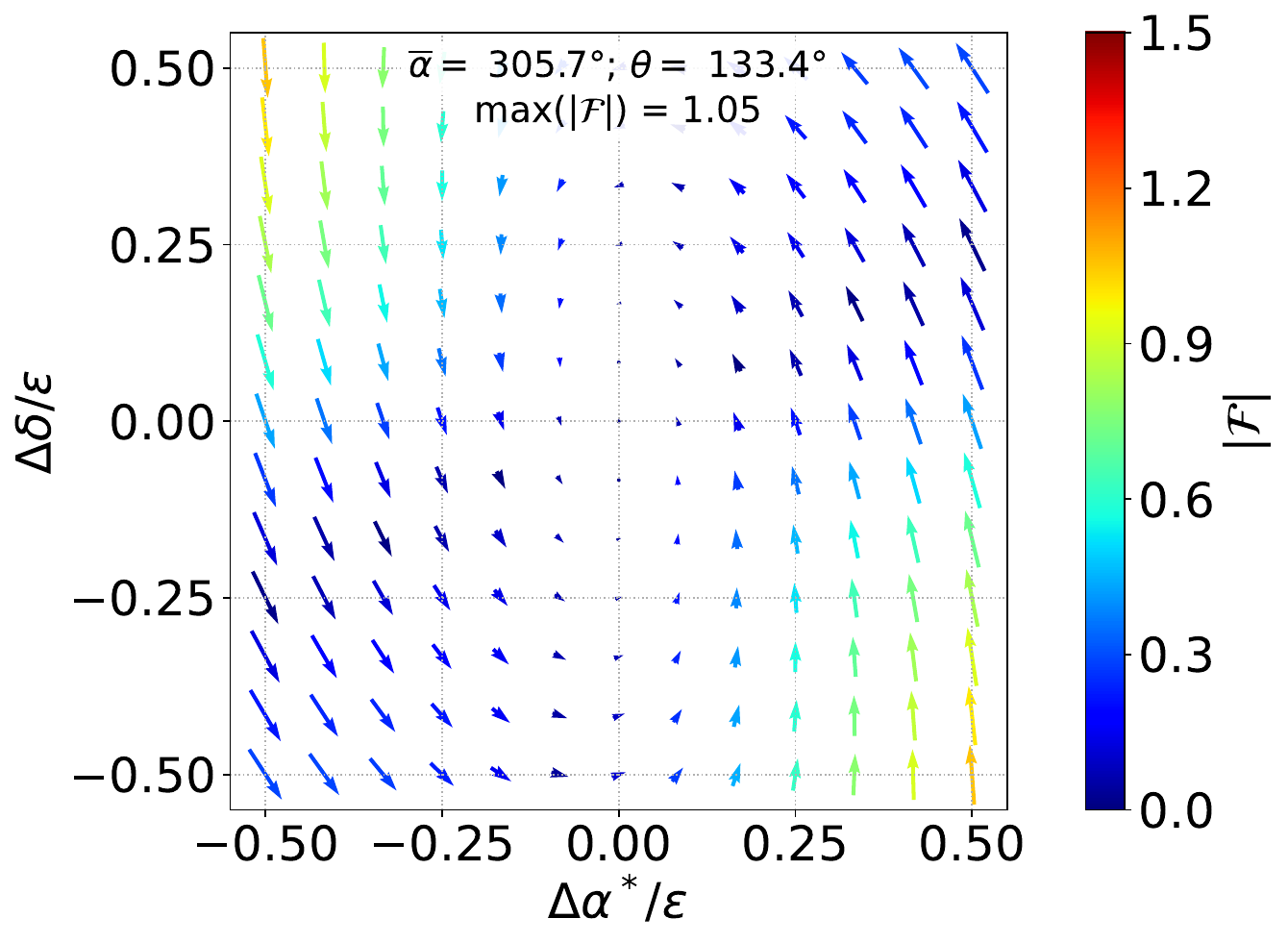}
	\caption{Vector field visualizations of the absolute (left
          column) and differential (right column) astrometric GW
          effects within an $(\varepsilon\times\varepsilon)$-sized FoV,
          for different positions on the sky relative to the GW
          propagation direction, as indicated at the top of each plot
          with the ($\overline{\alpha}, \theta$) of the center. In the
          left column, the absolute GW effect is plotted, i.e., the
          displacement of every point due to the GW, normalized to the
          maximum shift $\Delta_{\rm max}$ for this GW.  The gray
          arrows in this column are scaled independently to optimize
          visibility. At the top of each plot in this column, the
          maximum overall displacement, normalized to the maximal GW
          effect $\sqrt{(\delta\alpha^*)^2 + (\delta\delta)^2}$ in
          terms of $\Delta_{\rm max}$, is indicated. The right column
          shows the differential GW effect referenced to the central
          point with $(\Delta\alpha^*, \Delta\delta) = (0,0)$.  The
          arrow length in this column is determined by the difference
          between the displacement at the point and the displacement
          at the center.  The colors in the plots indicate the
          normalized angular change $|\mathcal{F}|$ with respect to
          the center point. The asterisk in $\Delta\alpha^*$ means that the difference in right ascension is a true arc, thus: $\Delta\alpha^* \equiv (\Delta\alpha) \cos \delta$. All plots have been created using the
          gnomonic projection. It should be noted that all arrow
          lengths are significantly exaggerated for illustrative
          purposes compared to typical GW signal magnitudes.  A
          detailed discussion can be found in the text.
	\label{fig__vector_field}}
\end{figure}

Figure~\ref{fig__vector_field} shows four examples of these
simulations. Each row of plots corresponds to a FoV area at a
different location on the sky. In this way, the variation of the
astrometric GW effect depending on the position can be seen. The
plots again show normalized values of both absolute (right column) and
differential (left column) GW effect.

A number of effects can be seen in Fig.~\ref{fig__vector_field}.
Generally, the absolute astrometric GW
effect inside a FoV appears basically uniform at first glance, as can be seen in the left
column. An exception is observed in the top-left panel of
Fig.~\ref{fig__vector_field}, where the FoV is precisely directed
towards the GW emitter ($\overline{\alpha} = 0^\circ;~\theta =
180^\circ$).

At this specific location, the overall GW-induced effect is minimal
(see e.g. Eq.~\ref{estimate1}), but lacks any dominant overall
shift. Consequently, the GW signal in this case is entirely differential, as evident
when compared to its corresponding plot in the right column. A
similar picture would be visible for the opposite point in the sky, in
the GW propagation direction ($\overline{\alpha} = 0^\circ;~\theta =
0^\circ$). Even a modest $5^\circ$ displacement of the FoV from the
GW emitter, as shown in the second row with ($\overline{\alpha} =
0^\circ;~\theta = 175^\circ$), results in all absolute positional changes
being overwhelmingly dominated by an average shift.  The differential
effect, however, is strongest there, inside one of the four petals of
the patterns discussed before. At other areas on the sky (two bottom
rows) the overall shift changes direction and the differential changes
are generally weaker. Notably, at the point ($\overline{\alpha} = 90^\circ;~\theta = 90^\circ$)
(row 3 in Fig.~\ref{fig__vector_field}), where the absolute GW effect
is maximal, the differential GW effect is significantly smaller compared to,
e.g., the case of ($\overline{\alpha} = 0^\circ;~\theta = 175^\circ$) (row 2 in Fig.~\ref{fig__vector_field}),
where the
maximum absolute GW effect is only 8\% of the maximal possible for this GW.  A further
effect is discernible in the bottom row, with the randomly selected
center position of ($\overline{\alpha} = 305.7^\circ;~\theta =
133.4^\circ$), where the differential GW effect in the FoV exhibits a
significant rotational component.  This leads to an interesting
consequence that the angular changes compared to the center are
negligible (note the relatively large dark blue arrows).
Angular separation between the center
and other sources would only change in the two corners of the FoV, where
a linear component in the differential shifts is present. Such
rotation-like and shear-like, differential patterns exist at many
positions on the sky.

We stress that the differential GW effects shown in the right column of
Fig.~\ref{fig__vector_field} are computed with respect to the central
point of the FoV. If the reference point is selected differently
(e.g. in one of the FoV's corners), the resulting plots can appear
significantly different, and the maximum absolute normalized angular
change, $\max(|\mathcal{F}|)$, may vary considerably.

\subsection{GW-induced astrometric signal and the plate solutions for small-field astrometry}

Standard plate corrections, like linear affine plate models
(e.g. translation, rotation, scale, and shear) or polynomial
corrections, must certainly be applied for small-field astrometric
observations. For high-accuracy astrometric solutions, the correction
parameters may even need to be time dependent.
Fig.~\ref{fig__vector_field} highlights a crucial finding in this
respect: the variation of the differential effect over the FoV is very
smooth, and even very simple plate corrections, like a
quadratic correction in each axis, would absorb most of the
differential effect in the FoV. In this way, the standard small-field
calibrations will most likely dramatically reduce the GW imprint on
the astrometric results.

To test this, we took the simulated differential GW effects from
Fig.~\ref{fig__vector_field} and applied a basic plate correction. We
fitted 2D bivariate Legendre polynomials up to order three for each
component of the shift (separately in both coordinates) and subtracted
the result from the differential effect shown in
Fig.~\ref{fig__vector_field}. The maximal differential effect for the
example with the highest differential magnitude, with
($\overline{\alpha} = 0^\circ;~\theta = 175^\circ$), decreased by a
factor of more than 20\,000, leaving a completely negligible residual
signal. For ($\overline{\alpha} = 90^\circ;~\theta = 90^\circ$) and
($\overline{\alpha} = 305.7^\circ;~\theta = 133.4^\circ$) the
differential GW signal is virtually completely absorbed by the
calibration polynomials. Even in the exact direction of the GW
emitter ($\overline{\alpha} = 0^\circ;~\theta = 180^\circ$) a simple
third-order correction attenuates the differential signal by more than
a factor of 2.

\section{Concluding remarks}
\label{sec__conclusion}

In this article, we investigated the effect of the astrometric GW
signal in small-field astrometry. The astrometric effect generated by
GWs will be tiny in any case. A detection of even the strongest
conceivable signals will require instruments at the absolute forefront
of technology. Practical detections of GWs with astrometry should
hence rely on long series of observations and a large number of
observed sources.  Given the relatively small FoV ($\varepsilon$) and
the correspondingly small number of sources observed by such
instruments, the prospects for detecting GWs with small-field astrometry
projects are rather bleak. For realistic FoVs (e.g., up to some
degrees in extent), the measurable effects introduced by GWs are tiny
compared to the magnitude of absolute GW effects, even over extended
periods of time. This is the fundamental difference from global
astrometry, such as \gaia\ \cite{2016A&A...595A...1G} and
\gaiaNIR\ \cite{2021ExA....51..783H}, where angular changes of pairs
of stars separated by a large angle (e.g. the basic angle of
$106.5^\circ$ for \gaia) are observed.  With such a large angular
distance, a significant number of observations will contain a signal
equivalent to the full magnitude of the absolute GW effect. Since
global scanning astrometry like \gaia\ observes the whole sky multiple
times, billions of observed objects and long duration of observations
can be used, increasing the chances of detecting GWs.

\section*{Acknowledgments}
This work is financially supported by ESA grant 4000115263/15/NL/IB
and the German Aerospace Agency (Deutsches Zentrum f\"ur Luft- und
Raumfahrt e.V., DLR) under grants 50QG1402 and 50QG2202. We also
thank the Center for Information Services and High Performance (ZIH)
at TU Dresden for providing a considerable amount of computing time.
The work contributed by L. Lindegren is supported by the Swedish National Space Agency.
Finally, we also thank Dr. Enrico Gerlach for fruitful discussions about some problems discussed here.

\section*{Data Availability}
The data that support the findings of Figs.~\ref{fig__non_histogram_theta} to \ref{fig__skymaps_and_non_histograms}, and Table~\ref{tab__stats_for_random} this article are openly available \cite{dataforthispaper}. Other data are available from the authors upon reasonable request.

\section*{Appendices}
\appendix
\section{Theoretical derivation of  astrometric effects of a plane gravitational wave}
\label{sec__appendix}

In this Appendix, we provide a detailed theoretical discussion of the calculation
of astrometric observable effects that are influenced by a gravitational wave.
As before, we restrict the discussion to a plane monochromatic GW.
In line with the notation used in Sect.~\ref{sec__max_change} we denote the observable angle
between two incident light rays originating from sources A and B
in the presence of a GW as $\psi_\mathrm{AB}^{\rm gw}$.
As is well known, this angle can be computed in different ways within
the framework of linear gravity used here.
A standard way is to project the null tangent vectors
to the two light rays at the moment of observation into the rest-space
of the observer in a suitably chosen coordinate system, with a
well-defined space-time metric tensor employed to compute $\psi_\mathrm{AB}^{\rm gw}$ by
means of scalar products (see below).
	
In the tetrad formalism, these null tangent vectors are projected onto
the co-moving tetrad system of the observer by means of the full
metric tensor, and $\psi_\mathrm{AB}^{\rm gw}$ is computed with the Euclidean metric in the
observer's 3-space.

If the observer has the possibility to operationally realize such
co-moving tetrads (e.g., by means of some mechanical structure), then
astrometric observables involving a single light ray can be defined (e.g., \cite{2025A&A...695A.172G}).

In addition, our derivations of the so-called {\it source terms} as well
as our treatment of a moving observer might be of general interest. To
the best of our knowledge, these issues have not been considered
elsewhere in such detail.

Below we use fairly standard notations. Greek indices $\alpha, \beta,
\dots$ running from 0 to 3 indicate all four space-time components of
the corresponding variable. Latin indices $a,b,\dots$ run from 1 to 3
and refer to three spatial components of the corresponding variable.
The Kronecker delta is $\delta^{ij}={\rm diag}(1,1,1)$. We use
Einstein's summation convention for both types of indices, independent
of the position of repeated indices: e.g., $x^i\,x^i\equiv
(x^1)^2+(x^2)^2+(x^3)^2$. A dot over any quantity designates the total
derivative with respect to the coordinate time of the corresponding
reference system: e.g., $\dot a=\displaystyle{da\over dt}$.  The
3-dimensional coordinate quantities (``3-vectors'') referred to the
spatial axes of the corresponding reference system are set in
boldface: $\ve{a}=a^i$. The absolute value (Euclidean norm) of a
``3-vector'' $\ve{a}$ is denoted as $|\ve{a}|$ and can be computed as
$|\ve{a}|=(a^1\,a^1+a^2\,a^2+a^3\,a^3)^{1/2}$.  The scalar product of
any two ``3-vectors'' $\ve{a}$ and $\ve{b}$ with respect to the
Euclidean metric $\delta_{ij}$ is denoted by $\ve{a}\,\cdot\,\ve{b}$
and can be computed as
$\ve{a}\,\cdot\,\ve{b}=\delta_{ij}\,a^i\,b^j=a^i\,b^i$.  The vector
product of any two ``3-vectors'' $\ve{a}$ and $\ve{b}$ is denoted
by $\ve{a}\times\ve{b}$ and can be computed as
$\left(\ve{a}\times\ve{b}\right)^i=\varepsilon_{ijk}\,a^j\,b^k$, where
$\varepsilon_{ijk}=(i-j)(j-k)(k-i)/2$ is the fully antisymmetric
Levi-Civita symbol.

\subsection{The metric tensor}\label{AppendixA}

The calculations will be performed in the framework of linear
gravity, using harmonic coordinates 
$x^{\mu} = \left(ct, x^1, x^2, x^3\right)$,
where, in the space-time region of interest, the metric tensor is of the form
\begin{eqnarray}
	g_{\alpha\beta} &=& \eta_{\alpha\beta} + h_{\alpha\beta}\,, \quad{\rm with} \quad |h_{\alpha\beta}| \ll 1 \, ,
	\label{metric_1}
\end{eqnarray}
 $\eta_{\alpha\beta} = {\rm diag}\left(-1, +1, +1, +1\right)$ and terms of order $\vert h_{\alpha\beta}\vert^2$ will be neglected.
 In (harmonic) TT-coordinates, where $h_{00} = h_{0i}= 0$, $h_{ij}$ is assumed  to be of the form
 \cite{2018CQGra..35d5005K,2025A&A...695A.172G}: 
\begin{eqnarray}
        h_{ij}\left(t,\ve{x}\right) &=& C_{ij}\,\cos\Phi + S_{ij}\,\sin\Phi\,,  
        \label{metric_4}
\end{eqnarray}
\noindent
with the phase
\begin{eqnarray}
        \Phi &=& \frac{2 \pi \nu}{c} \left(c t - \ve{p} \cdot \ve{x}\right) \, ,
        \label{phase}
\end{eqnarray}
where $\nu$ denotes the frequency of the GW and $\ve{p}$ the Euclidean
unit vector ($p^i p^i = 1$) in the GW's propagation direction. The
tensorial coefficients in \eqref{metric_4} are given by $C_{ij} =
p_{ij}^{+} h_c^{+} + p_{ij}^{\times} h_c^{\times}$ and $S_{i j} =
p_{ij}^{+} h_s^{+} + p_{ij}^{\times} h_s^{\times}$, where $h_c^{+}$,
$h_s^{+}$, $h_c^{\times}$, $h_s^{\times}$ are four independent strain
parameters.  The matrices $p_{ij}^+$ and $p_{ij}^{\times}$ can be
written in the form
\begin{eqnarray}
	p_{ij}^{+} &=& \left(\bf{P}\,\bf{e}^{+}\,\bf{P}^{\rm T}\right)_{ij} \quad {\rm and} \quad  
	p_{ij}^{\times} = \left(\bf{P}\,\bf{e}^{\times}\,\bf{P}^{\rm T}\right)_{ij}
	\label{metric_p}
\end{eqnarray}
\noindent
with 
\begin{eqnarray}
        {\bf e}^{+}_{ij} &=&    
        \left(\begin{array}{ccc}
                + 1 & 0   & \;0 \\
                0   & - 1 & \;0 \\
                0   & 0   & \;0 \\
        \end{array}\right) \quad {\rm and} \quad 
        {\bf e}^{\times}_{ij} =  
        \left(\begin{array}{ccc}
                0 & + 1 & \;0 \\
                + 1 & 0 & \;0 \\
                0 & 0 & \;0 \\
        \end{array} \right) \, .
        \label{GW_1_2}
\end{eqnarray}
\noindent
Here ${\bf P}$ is the rotational matrix between the reference system
in which the gravitational wave propagates in the z-direction and the
coordinate system in which the direction of gravitational wave
propagation is $\ve{p}$
\cite{2018CQGra..35d5005K,2025A&A...695A.172G}. The $+$ and $\times$
parts of $h_{ij}$ correspond to the two polarization modes of the GW.

 \subsection{The geodesic equation and the null condition}
In the following, we will first consider  a single light ray emitted at the event $(t_0, \ve{x}_0)$
by some light source (star).

The geodesic equation and the null condition ($ds^2 = 0$) for light rays 
in linear gravity have been given in the literature,
e.g., \cite{Brumberg1991,1999PhRvD..59h4023K,1999PhRvD..60l4002K,2003A&A...410.1063K}.
In TT-gauge, using coordinate time $t$ as parameter, they take the form
\begin{eqnarray}
	\frac{\ddot{x}^i \left(t\right)}{c^2} &=& - h_{ij,0}\,\mu^j + \frac{1}{2}\,h_{jk,i}\,\mu^j \mu^k - h_{ij,k}\,\mu^j \mu^k 
	+ \frac{1}{2}\,h_{jk,0}\,\mu^i \mu^j \mu^k\,, 
\label{Geodesic_Equation_1PM}
\\
\frac{\left|\dot{\ve{x}}\left(t\right)\right|}{c} &=& 1 - \frac{1}{2}\,h_{i j}\,\mu^i \mu^j\,,
\label{null_condition_1PM}
\end{eqnarray}
\noindent
where the dot indicates the time derivative, a comma indicates a partial derivative ($f_{,i} \equiv \partial f/ \partial x^i$ and
$f_{,0} \equiv c^{-1}\,\partial f/ \partial t$) and 
\begin{eqnarray}
        \ve{\mu} &=& \frac{\dot{\ve{x}}\left(t\right)}{\left| \dot{\ve{x}}\left(t\right) \right|} \Bigg|_{t = t_0}
        \label{initial_condition_1}
\end{eqnarray}
\noindent 
is an Euclidean unit vector ($\mu^i \mu^i = 1$) that points in the spatial coordinate direction of the light ray at the moment of emission.

The solution of the homogeneous equation, $\ddot{x}^{i}\left(t\right)
= 0$, is given by the 'unperturbed light ray' (e.g., Eq.~(C24) in
\cite{2003A&A...410.1063K}),
\begin{eqnarray}
        \ve{x}_{\rm N}\left(t\right) = \ve{x}_0 + c \left(t - t_0\right) \ve{\mu}\, .
        \label{unperturbed_lightray}
\end{eqnarray}
\noindent
Note, that the Euclidean 'tangent vector' $\ve{\mu}$ is a free parameter
so far. It will be chosen later so that the perturbed light ray goes
through the event of observation $(t_1,\ve{x}_1)$.
Formally, this choice then shows first-order terms explicitly

\subsubsection{The first integration of geodesic equation} 

The function $\dot x^i(t)$ is obtained by integrating the geodesic
equation~\eqref{Geodesic_Equation_1PM} over the time coordinate from
$t_0$ to $t > t_0$ along the unperturbed light ray, i.e., by writing
$\vec{x} = \vec{x}_{\rm N}(t)$ in \eqref{metric_4} and in particular in the phase $\Phi$, which takes the form
\begin{eqnarray}
  \Phi(t)&=&\Phi_{\rm N}(t)=2\pi\nu\left(t - c^{-1}\,\ve{p} \cdot \ve{x}_{\rm N}(t)\right)
  \nonumber\\[3pt]
  &=&\Phi_{\rm N}(t_0)+2\pi\nu\,(1-\ve{p}\cdot\ve{\mu})\,(t-t_0)\,.
  \label{PhiN(t)=}
\end{eqnarray}  
In the special case where $\ve{\mu} = \ve{p}$, the phase $\Phi(t)$ is
constant along the light ray, and the gravitational wave, due to its
transversal character, does not influence the propagation of the
light-ray. In the following, we will assume that $\ve{\mu}\neq\ve{p}$,
but the limit $\ve{\mu}\to\ve{p}$ is discussed after
Eq.~\eqref{overline-h} below. The integrands are then pure functions
of $t$ and one gets
\begin{eqnarray}
  \frac{\dot{x}^i(t)}{c} &=& \frac{\dot{x}^i(t_0)}{c} 
	+ \frac{\Delta\dot{x}^i(t,t_0)}{c}\,,
	\label{first_integration}
\end{eqnarray}
\noindent
where
\begin{eqnarray}
	\frac{\dot{x}^i(t_0)}{c}  &=& 
\mu^i - \frac{1}{2}\,h_{jk}(t_0,\ve{x}_0)\,\mu^j\,\mu^k\,\mu^i
\end{eqnarray}
that follows from the null condition \eqref{null_condition_1PM} and
\begin{eqnarray}
\frac{\Delta\dot{x}^i(t,t_0)}{c} &=& c\,\int\limits_{t_0}^{t} dt\,\frac{\ddot{x}^i\left(t\right)}{c^2}
        \label{first_integration_4} 
        = \frac{\Delta\dot{x}^i(t)}{c} - \frac{\Delta\dot{x}^i(t_0)}{c}
        \label{lightray_11}
\end{eqnarray}
\noindent
and (here and below $\ve{x}_{\rm N}=\ve{x}_{\rm N}(t)$ with the corresponding time argument)
\begin{eqnarray}
	\frac{\Delta\dot{x}^i(t)}{c} &=& 
	+ \frac{1}{2}\,h_{jk}(t,\ve{x}_{\rm N})\,\mu^j\,\mu^k\,\frac{\mu^i - p^i}{1 - \ve{p}\cdot\ve{\mu}} - h_{ij}(t,\ve{x}_{\rm N})\,\mu^j\,.
	\label{lightray_12}
\end{eqnarray}

\subsubsection{The second integration of geodesic equation}

Considering the result \eqref{first_integration} and the initial condition $x^i\left(t_0\right)=x_0^i$ the second integration leads to
\begin{eqnarray}
	x^i\left(t\right) &=& x_0^i + c \left(t - t_0\right) \mu^i - \frac{1}{2} c \left(t - t_0\right) h_{jk}(t_0,\ve{x}_0)\,\mu^j\,\mu^k\,\mu^i 
	+ \Delta x^i\left(t,t_0\right) \,,
        \label{second_integration_5}
\end{eqnarray}
\noindent
where
\begin{eqnarray}
  \Delta x^i\left(t,t_0\right) &=& c\,\int\limits_{t_0}^{t} dt\,
  \left( \frac{\Delta\dot{x}^i\left(t\right)}{c} -  \frac{\Delta \dot x^i(t_0)}{c}\right) 
     = \Delta x^i\left(t\right) - \Delta x^i\left(t_0\right) - c \left(t - t_0\right) \frac{\Delta\dot{x}^i\left(t_0\right)}{c}\,. 
        \label{second_integration}
\end{eqnarray}

\noindent 
The term $\Delta x^i(t)$ can formally be written in the form
\begin{eqnarray}
	\Delta x^i\left(t\right) &=& - \frac{\lambda}{2\pi}\,{1\over1 - \ve{p}\cdot\ve{\mu}}\,
        \left[\,\frac{1}{2}\,{\overline h}_{jk}(t,\ve{x}_{\rm N})\,\mu^j\,\mu^k\,\frac{\mu^i - p^i}{1 - \ve{p}\cdot\ve{\mu}}
          - {\overline h}_{ij}(t,\ve{x}_{\rm N})\,\mu^j\,\right] \,, 
        \label{second_integration_t}
        \\
        {\overline h}_{ij}(t,\ve{x}_{\rm N})&=&-C_{ij}\,\sin\Phi_{\rm N} + S_{ij}\,\cos\Phi_{\rm N}\,.
        \label{overline-h}
\end{eqnarray}
\noindent
Here $\lambda=c/\nu$ is the wavelength of the GW. We note that, by using
$C_{ij} p^j=S_{ij} p^j=0$, one can see that $\Delta\dot{x}^i(t)$ for
any $t$ in Eq.~\eqref{lightray_12} tends to zero in the case
$\ve{\mu}\to\ve{p}$ even though the denominator with $1 -
\ve{p}\cdot\ve{\mu}$ itself tends to zero. The situation is trickier for
$\Delta x^i\left(t\right)$ in Eq.~\eqref{second_integration_t} which
diverges for $\ve{\mu}\to\ve{p}$. However, as one can see from
Eq.~\eqref{PhiN(t)=} the phase difference $\Phi_{\rm N}(t)-\Phi_{\rm
  N}(t_0)$ for a given $2\pi\nu\,(t-t_0)$ tends to zero when 
$\ve{\mu}\to\ve{p}$. This allows one to see that $\Delta
x^i\left(t\right) - \Delta x^i\left(t_0\right)$ in
Eq.~\eqref{second_integration} tends to zero when $\ve{\mu}$ tends to
$\ve{p}$.

\subsection{The boundary value problem}

In the previous Sections, we have solved the initial value problem for a single light ray that is emitted from $\ve{x}_0$, at coordinate time $t_0$, in a direction given by the Euclidean 3-vector $\ve{\mu}$ from \eqref{initial_condition_1}.
Now we imagine an observer that observes this light ray at coordinate position $\vx_1$ at some coordinate time $t_1$:
\begin{eqnarray}
	\ve{x}_1 &=& \ve{x}\left(t\right)\bigg|_{t = t_1}\, .
        \label{boundary_1}
\end{eqnarray}
\noindent
We note that while $\vx_1$ can be chosen arbitrarily and, given $\vx_0$
and $t_0$, defines $\ve{\mu}$, the moment of time $t_1$ is itself defined
by $\vx_0$, $\vx_1$, $\ve{\mu}$, $t_0$ and the metric tensor.

Inserting \eqref{boundary_1} into the equation for the light
trajectory \eqref{second_integration_5}, after some rewriting, leads
to the relation
\begin{eqnarray}
	\ve{\mu} &=& \ve{k} - \frac{1}{R}\,\ve{k} \times \Bigl(\bigl[\Delta \ve{x}(t_1)-\Delta \ve{x}(t_0)\bigr] \times \ve{k}\Bigr) 
	+ \ve{k} \times \left(\frac{\Delta\dot{\ve{x}}(t_0)}{c}\times \ve{k}\right)
	\label{transformation_k_n} \, ,
\end{eqnarray}
\noindent
where
\begin{eqnarray}
        \ve{k} &=& \frac{\ve{x}_1 - \ve{x}_0}{|\ve{x}_1 - \ve{x}_0|} 
        \label{vector_k}
\end{eqnarray}
\noindent
is the Euclidean spatial unit vector ($k^i k^i = 1$) that points from the emission point to the point of observation, and
$R = \vert \vx_1 - \vx_0 \vert$ is the Euclidean spatial coordinate distance between
the light source and the observer.

By inserting \eqref{transformation_k_n} into \eqref{first_integration} one obtains
\begin{eqnarray}
        \frac{\dot{\ve{x}}(t_1)}{c} &=& \ve{k} + \frac{\Delta\dot{\ve{x}}(t_1)}{c}
        - \frac{1}{R}\,\ve{k} \times \Bigl(\bigl[\Delta \ve{x}(t_1)-\Delta \ve{x}(t_0)\bigr] \times \ve{k}\Bigr)\,.
        \label{first_integration_5}
\end{eqnarray}
\noindent
The part of the $1/R$-term with $\Delta \ve{x}(t_0)$ depends on the
GW-field at the source and is often called ``source term''. Effectively,
the terms with $\Delta\ve{x}$ in Eq.~\eqref{first_integration_5} are
proportional to $\lambda/R$. Considering GWs with periods $P_{\rm
  GW}=1/\nu<30$ years (see \cite{2025A&A...695A.172G} for a
discussion) we have $\lambda<9.2$\,pc.  Except for observations of the
relatively small number of nearby stars within a distance of e.g.
100\,pc, we have $\lambda/R\lesssim0.1$.

We see two lines of argument for why the $\lambda/R$ terms can be
neglected. First, normally we are interested in the case where a large
number of astrometric sources are observed. In this case, the terms
proportional to $\lambda/R$ are not correlated with each other for
different stars. Moreover, for most of the stars these terms cannot be
computed with sufficient accuracy, because the distances are not
known so precisely, even in the Gaia era. Therefore, as is also
often argued in the literature, those terms can be considered as an
additional stochastic noise in the data. The second line of argument is
related to the magnitude of the $\lambda/R$ terms for individual
observations. For $\lambda/R\lesssim0.1$ one can see that the term
$\Delta\dot{\ve{x}}(t)$ dominates the signal as soon as the overall GW
effect is comparable to $\Delta_{\rm max}$ introduced in
\cite{2025A&A...695A.172G}. The $\lambda/R$ terms can become
comparable to, or even larger than, the effect of
$\Delta\dot{\ve{x}}(t)$ only in cases where the overall GW effect is
negligibly small. However, if astrometric observations of nearby stars
such as the $\alpha$ Centauri system are used, the $\lambda/R$ term may
be of interest.

In the following we will neglect this term and continue with the
following first-order expression:
\begin{eqnarray}
        \frac{\dot{\ve{x}}\left(t_1\right)}{c} &=& \ve{k} + \frac{\Delta\dot{\ve{x}}\left(t_1\right)}{c}\,,
        \label{first_integration_6}
\end{eqnarray}
\noindent
where
\begin{eqnarray}
        \frac{\Delta\dot{x}^i\left(t_1\right)}{c} &=&
        + \frac{1}{2}\,h_{jk}\left(t_1,\ve{x}_1\right)\,k^j\,k^k\,\frac{k^i - p^i}{1 - \ve{p}\cdot\ve{k}} - h_{ij}\left(t_1,\ve{x}_1\right)\,k^j \, .
        \label{observer_term_k}
\end{eqnarray}

\subsection{The worldline of an observer}
\label{AppendixE}
In any realistic observational setup (like e.g., \gaia or \gaiaNIR) the
observer (satellite) will undergo a complex motion described by
some ephemeris in a suitably chosen coordinate system. Because of
aberration, the problem of a moving observer is by no means academic
for the central problem discussed here. Accordingly, we
consider an observer in TT coordinates with worldline $x^{\mu}_{\rm
  obs}\left(\tau\right)$, where $\tau$ is the observer's proper time
related to the fundamental length element $ds$ along their
worldline. For the construction of astrometric observables for such an
observer, one needs the observer's (normalized) four-velocity
\begin{eqnarray}
  u^{\mu} &=& {1\over c}\,\frac{dx^{\mu}_{\rm obs}(\tau)}{d\tau} 
\end{eqnarray}
\noindent
with $g_{\mu\nu}\,u^{\mu} \,u^{\nu} = - 1$.  Effects of the observer's
motion will now be considered with $\vec{v}$ being the observer's TT
coordinate velocity. The normalized
4-velocity is then given by
\begin{eqnarray}
        u^\alpha &=& \gamma_h\,(+1\,,\beta^i)\,,
        \label{four_velocity}
\end{eqnarray}
\noindent
where $\beta^i = v^i/c$, $v^i=dx^i_{\rm obs}/dt$ is the coordinate velocity of observer, and
\begin{eqnarray}
        \gamma_h &=& (1 - \vec{\beta}^2 - h_{ij} \beta^i \beta^j)^{-1/2} \, .
        \label{gamma_h}
\end{eqnarray}

\subsection{The observed angle between two incident light rays}\label{sec__apx_angle}

The well-known formula for the angle $\psi_\mathrm{AB}^{\rm gw}$ between two incident light rays ($A$ and $B$) as measured by an observer reads:
\begin{eqnarray}
	\cos \psi_\mathrm{AB}^{\rm gw} &=& g_{\alpha \beta}\,\frac{\lbar_A^{\,\alpha}}{\left|\lbar_A\right|}\,\frac{\lbar_B^{\,\beta}}{\left|\lbar_B\right|}\,, 
	\label{angle_1}
\end{eqnarray}
\noindent
where the quantities of the right-hand side refer to the event of
observation $(t_1, \vx_1)$.  $\lbar^{\,\alpha}$ is the tangent null vector to the
light trajectory $A$ or $B$, projected into the rest-space of the
observer and
\begin{equation}
  \left|\,\lbar\,\right| = \left(g_{\mu\nu}\,\lbar^{\,\mu}\, \lbar^{\,\nu}\right)^{1/2} \,.
  \label{|lbar|}
\end{equation}
\noindent
For our purposes it is sufficient to parametrize a light trajectory
with coordinate time $t$. Then, the null tangent vector to a light ray
takes the form
\begin{eqnarray}
        l^{\alpha} &=& {1\over c}\,\frac{d x^{\alpha}}{dt} 
        = \left(1,\frac{\dot{\ve{x}}\left(t\right)}{c}\right), 
        \label{lightray_1}
\end{eqnarray}
\noindent
where $\dot{\ve{x}}\left(t\right)$ is given by \eqref{first_integration_6}.

The projected tangent vector $\lbar^{\,\alpha}$ is then given by 
\begin{eqnarray}
	\lbar^{\,\alpha} &=& P^{\alpha}_{\beta}\,l^{\beta}\,, 
	\label{lightray_2}
\end{eqnarray}
\noindent
where 
\begin{eqnarray}
P^{\alpha}_{\beta} = \delta^{\alpha}_{\beta} + g_{\beta\gamma} u^{\alpha}\,u^{\gamma}\, 
\label{projector}
\end{eqnarray}
\noindent
projects vectors into the rest-space of the observer orthogonal to his four velocity $u^\alpha$.
Then
\begin{eqnarray}
        \lbar^{\,\alpha} &=& l^{\alpha} - \cE\,u^{\alpha}\,,
        \label{tangent_vector_moving_observer}
\end{eqnarray}
\noindent
where
\begin{eqnarray}
        \cE &=& - g_{\mu\nu}\,l^{\mu}\,u^{\nu} = \left| l^{\mu}\,u_{\mu} \right|\,.
        \label{abbreviation_tangent_vector_moving_observer}
\end{eqnarray}
\noindent
The four-vector in \eqref{tangent_vector_moving_observer} is space-like 
and one finds for the norm
\begin{eqnarray}
        \left|\,\lbar\,\right| &=& \cE = \gamma_h \, \left( 1 - \ve{k} \cdot \ve{\beta} - \frac 12 h_{jk}
        k^j k^k \, \frac{\ve{\beta} \cdot \ve{k} - \ve{\beta}\cdot \ve{p}} {1 - \ve{p}\cdot \ve{k}}\right)\,.
        \label{lightray_norm}
\end{eqnarray}
\noindent
We finally obtain $(\eta = 1/\cE)$
\begin{eqnarray}
	\frac{\bar{l}^{\,\alpha}}{\left|\,\lbar\,\right|} &=& \left( \eta - \gamma_h, \eta \,\frac{\dot \vx}{c} - \gamma_h \ve{\beta} \right)\,.
\end{eqnarray}
\noindent
With this expression, applied to the two light rays, $A$ and $B$, and
the scalar product taken with the space-time metric as shown in Eq.~\eqref{angle_1} one obtains 
$\cos \psi_\mathrm{AB}^{\rm gw}$ as seen by the observer with coordinate velocity $v^i$ and
located at $\vx_1$ at $t = t_1$. As we have neglected the source terms
in the right-hand side of \eqref{angle_1} is completely determined by
the event of observation.

To compare our formulation with those existing already in the
literature we derive here the explicit formula for the case of an
observer at rest in our coordinate system, where $\ve{\beta} = 0$ and
$\gamma_h = 1$, so that the observer's four-velocity reads $u^{\mu} =
\left(1,0,0,0\right)$. For this special case we get $\cE = 1$ and
\begin{eqnarray}
        \frac{\lbar^{\,i}}{\left|\,\lbar\,\right|} &=& k^i + \frac{1}{2}\,h_{jk}\,k^j\,k^k\,\frac{k^i - p^i}{1 - \ve{p}\cdot\ve{k}} - h_{ij}\,k^j\,.
        \label{unit_tangent_vector_observer_at_rest}
\end{eqnarray}
\noindent
Note, that the time-component vanishes here ($\lbar^{\,0} = 0$).  By
using this relation in Eq.~\eqref{angle_1} for our two light rays $A$
and $B$, we get our final result for the observed angle between these
two light rays in the form:
\begin{eqnarray}
	\cos \psi_\mathrm{AB}^{\rm gw} &=& \ve{k}_A \cdot \ve{k}_B 
	+ \frac{1}{2}\,h_{jk} \bigg[k_A^j\,k_A^k\,\frac{\ve{k}_A \cdot \ve{k}_B - \ve{k}_B \cdot \ve{p}}{1 - \ve{k}_A \cdot \ve{p}} 
	+ k_B^j\,k_B^k\,\frac{\ve{k}_A \cdot \ve{k}_B - \ve{k}_A \cdot \ve{p}}{1 - \ve{k}_B \cdot \ve{p}}\bigg]  
	- h_{ij}\,k_A^i\,k_B^j\,,
	\label{final_result_angle}  
\end{eqnarray}
\noindent
where the right-hand side refers to the event of observation. From
\eqref{final_result_angle} one sees that in the limit when two sources $A$ and $B$ get
closer and closer to each other one has
\begin{equation}
        \lim_{A \rightarrow B}\;\cos \psi_\mathrm{AB}^{\rm gw} = \lim_{A \rightarrow B}\;\ve{k}_A \cdot \ve{k}_B = 1
        \label{limit_A_B}
\end{equation}
\noindent
as one can expect from a continuous vector field of the astrometric GW
signal
\cite{Article_Book_Flanagan,2018CQGra..35d5005K,2025A&A...695A.172G}. One
can see that the same limit holds true also for a moving observer.

\subsection{The observed angle $\psi_\mathrm{AB}^{\rm gw}$ in the tetrad formalism} 

The angle $\psi_\mathrm{AB}^{\rm gw}$ in Eq.~\eqref{angle_1} represents an observable that,
theoretically, is described by a scalar, i.e., a coordinate independent quantity.
The quantities on the right-hand side of \eqref{angle_1} have been
expressed in terms of tensor components with respect to the TT
coordinates at the event of observation.  

In the context of our problem, local proper coordinates are often
employed that have a direct physical meaning.
Such coordinates are usually constructed as tetrad-induced
quantities, e.g., \cite{MTW,1989racm.book.....S,1991PhRvD..43.3273D}.
These tetrads form a set of four orthonormal basis vectors,
$e^\mu_{(\alpha)}$, that act as tangent vectors to the local coordinate lines.
The indices $(\alpha)$ label the tetrad components, while the indices $\mu$ are tensor indices. 
These tetrads are defined along the worldline $x^{\mu}_{\rm obs}\left(\tau\right)$
of the observer (defining the origin of local coordinates),
and obey the orthonormality condition (e.g.,
\cite{MTW,Brumberg1991,2011rcms.book.....K} and more
specifically \cite{1992AJ....104..897K,2004PhRvD..69l4001K}): 
\begin{eqnarray}
 g_{\mu\nu}\,e^{\mu}_{(\alpha)}\,e^{\nu}_{(\beta)}  &=& \eta_{\alpha\beta}\,.
\label{tetrad_formalism_5}
\end{eqnarray}
\noindent 
The zeroth time-like tetrad vector $e^\mu_{(0)}$ is chosen as normalized 4-velocity of the observer 
\begin{eqnarray}
 	e^{\mu}_{(0)} &=& u^{\mu}\,,
 	\label{tetrad_formalism_10} 
 \end{eqnarray}
\noindent 
so that the vectors $e^\mu_{(i)}, i = 1,2,3$ span the 3-space of the
observer at a certain event on the worldline $x^{\mu}_{\rm
  obs}\left(\tau\right)$ of the observer.

For an arbitrary vector $A^\mu$ on the observer's worldline, we can then write
\begin{equation}
	A^\mu = A^{(\alpha)} e^{\mu}_{(\alpha)}
\end{equation}
and
\begin{equation}
	g_{\mu\nu} A^\mu e^\nu_{(\beta)} = \eta_{\alpha\beta} A^{(\alpha)} \, .
\end{equation}
Considering two such vectors $A^\mu$ and $B^\nu$ one gets
\begin{equation}
	g_{\mu\nu} A^\mu B^\nu = \eta_{\alpha\beta} A^{(\alpha)} B^{(\beta)} \, .
\end{equation}

We may assume that in high-precision astrometric satellite missions
the ``observer'' has the possibility to operationally realize such
space-like basic vectors (by means of ``quasi-rigid'' mechanical
structures). The tetrad formalism can be used to compute the
observables in astrometry in several different ways (see an overview
in \cite{2004PhRvD..69l4001K}). Here, we prefer the following approach.
For a tangent vector to some light ray projected into the rest space
of an observer, $\lbar^\alpha$ as in Eq.~\eqref{lightray_2}, its
tetrad components are $\lbar^{(0)} = 0$ and $\lbar^{(i)} = l^{(i)}$ (since
$l^{(i)}$ lies in in the observer's instantaneous 3-space). The
spatial components $l^{(i)}$ can be considered as components of an
Euclidean 3-vector $\ve{l}$ with Euclidean norm $\vert\,\ve{l}\,\vert
= (l^{(i)} l^{(i)})^{1/2}$. We note also that $\vert\,\ve{l}\,\vert=
\left|\,\lbar\,\right|$, where $\left|\,\lbar\,\right|$ is defined by
Eqs.~\eqref{|lbar|} and \eqref{lightray_norm}.  Then the Cartesian
components $l^{(i)}$ of a (null) tangent vector $l^\alpha$ to some
incident light ray are observable (coordinate-independent quantities).
The corresponding (negative) Euclidean unit vector
\begin{equation}
s^i = - \frac{l^{(i)}} {\vert\,\ve{l}\,\vert} 
\end{equation}
\noindent
then has observable components that can be formulated as directional
angles $(\alpha,\delta)$ of a single light ray towards the astrometric
source as seen by the observer, i.e., ${\bf s} = (\cos \alpha \cos
\delta, \sin \alpha \cos \delta, \sin \delta)$ (e.g.,
\cite{2025A&A...695A.172G}). The observed angle between two incoming
light rays (or equivalently between two observed directions $\ve{s}_A$
and $\ve{s}_B$) can then be computed as
\begin{eqnarray}
	\cos \psi_\mathrm{AB}^{\rm gw} &=&  \delta_{ij}\, \frac{l_A^{(i)}\,l_B^{(j)}}{\vert\, \ve{l}_A \vert \, \vert\, \ve{l}_B \vert}\,,
	\label{angle_B}
\end{eqnarray}
\noindent
where the right-hand side has to be taken at the event of
observation. Eq.~\eqref{angle_B} is equivalent to Eq.~\eqref{angle_1},
but is written using the tetrad components of the corresponding
vectors.

For an observer at rest in TT coordinates, the tetrad vectors are
given by \cite{2004PhRvD..69l4001K,Brumberg1991}
\begin{equation}
e^{0}_{(0)} = 1\,,\ \ 
        e^{i}_{(0)} = 0\,,\ \ 
        e^{0}_{(i)} = 0\,,\ \
        e^{i}_{(j)} = \delta_{ij} - \frac{1}{2}\,h_{ij}\label{tetrads} 
\end{equation}
\noindent 
and one finds the components of $s^i$ in the form
\begin{eqnarray}
-s^i={l^{(i)}\over {\vert\,\ve{l}\,\vert}} 
&=& 
k^i + \frac{1}{2}\,h_{jk}\left(t_1,\ve{x}_1\right)\,k^j\,k^k\,\frac{k^i - p^i}{1 - \ve{p}\cdot\ve{k}} - \frac{1}{2}\,h_{ij}\left(t_1,\ve{x}_1\right)\,k^j  
        \label{proof_tetrad_observer_i} \, .
\end{eqnarray}
\noindent 
This result is in agreement with Eq.~(58) in
\cite{Article_Book_Flanagan} as well as Eq.~(A1) in
\cite{2025A&A...695A.172G}. Our final result
\eqref{final_result_angle} for $\cos \psi_\mathrm{AB}^{\rm gw}$ can then be recovered by
substituting Eq.~\eqref{proof_tetrad_observer_i} into
Eq.~\eqref{angle_B}.

Eq.~\eqref{angle_B} is valid for a moving observer as well. However, the
tetrad for a moving observer is more complicated and related to
\eqref{tetrads} by a Lorentz boost, as discussed in
\cite{2004PhRvD..69l4001K}.

We note that in \cite{MT_Crosta,2025NatSR..1532908S} obviously the tetrad components of
\eqref{proof_tetrad_observer_i} were used in expression
\eqref{angle_1} instead of \eqref{angle_B}. This fatal flaw eventually
leads to their incorrect final result for $\cos \psi_\mathrm{AB}^{\rm gw}$.
Another discussion of this flaw can be found in \cite{2025arXiv250718593V}.

\bibliography{reports}

\newpage

\end{document}